\documentclass[letterpaper]{sae}
\usepackage{color}
\usepackage{fancyhdr}
\usepackage{lastpage}
\usepackage{etoolbox}

\PaperTitle{Engine and Aftertreatment Co-Optimization of Connected HEVs via Multi-Range Vehicle Speed Planning and Prediction} 
\AddAuthor{Qiuhao Hu, Mohammad Reza Amini, Yiheng Feng, Zhen Yang, Hao Wang, Ilya Kolmanovsky, and Jing Sun}{University of Michigan, Ann Arbor, MI, USA}
\AddAuthor{Ashley Wiese, Zeng Qiu, and  Julia H. Buckland Seeds}{Ford Motor Company, Dearborn, MI, USA}
\PaperNumber{2020-01-0590}
\usepackage[utf8]{inputenc}
\usepackage[english]{babel}
\usepackage{graphicx}
\usepackage[utf8]{inputenc}
\usepackage{algorithm}
\usepackage[noend]{algpseudocode}
\usepackage{amsmath}
\usepackage{array}
\usepackage{booktabs}
\usepackage{nomencl}
\usepackage{rotating}
\usepackage{indentfirst}
\usepackage{amsmath}
\usepackage{amssymb}
\usepackage{hyperref}
\usepackage{setspace}
\usepackage{extdash}
\usepackage{graphicx}
\usepackage{graphics}
\usepackage{epstopdf}
\usepackage{lipsum}
\usepackage{multirow}
\usepackage{tablefootnote}
\usepackage[usenames,dvipsnames]{xcolor}
\usepackage{cite}
\usepackage{mathtools}
\usepackage{mathrsfs}

\begin{document}
\graphicspath{{figures/}}
%\doublespace

\pagestyle{fancy}
\fancyhf{}
\renewcommand{\headrulewidth}{0pt}
\fancypagestyle{firstpage}{\lhead{\vspace{-0.75 cm} \small \textbf{{\fontfamily{cmss}\selectfont
Citation: \textcolor{blue}{Hu, Q., Amini, M.R., Feng, Y., Yang, Z., Wang, H., Kolmanovsky, I., Sun, J., Wiese, A., Qiu, Z., and Buckland, J., ``Engine and Aftertreatment Co-Optimization of Connected HEVs via Multi-Range Vehicle Speed Planning and Prediction,'' \\SAE Technical Paper 2020-01-0590, 2020.}}}}}
\fancyfoot[L]{Page \thepage \hspace{1pt} of \pageref{LastPage}} 

\maketitle 

\thispagestyle{firstpage}%<------------------

\AtEndEnvironment{algorithm}{\kern2pt\hrule\relax\vskip3pt\@algcomment}

\newcommand\algcomment[1]{\def\@algcomment{\footnotesize#1}}

%\pagestyle{fancy}
%\fancyhf{}
%\renewcommand{\headrulewidth}{0pt} 
%\fancyfoot[L]{Page \thepage \hspace{1pt} of \pageref{LastPage}} 

\section{Abstract} \vspace{-8pt}
Connected vehicles (CVs) have situational awareness that can be exploited for control and optimization of the powertrain system. While extensive studies have been carried out for energy efficiency improvement of CVs via eco-driving and planning, the implication of such technologies on the thermal responses of CVs (including those of the engine and aftertreatment systems) has not been fully investigated. One of the key challenges in leveraging connectivity for optimization-based thermal management of CVs is the relatively slow thermal dynamics, which necessitate the use of a long prediction horizon to achieve the best performance. Long-term prediction of the CV speed, unlike the short-range prediction based on vehicle-to-infrastructure (V2I) and vehicle-to-vehicle (V2V) communications-based information, is difficult and error-prone.

\vspace{-4pt}
The multiple timescales inherent to power and thermal systems call for a variable timescale optimization framework with access to short- and long-term vehicle speed preview. To this end, a model predictive controller (MPC) with a multi-range speed preview for integrated power and thermal management (iPTM) of connected hybrid electric vehicles (HEVs) is presented in this paper. The MPC is formulated to manage the power-split between the engine and the battery while enforcing the power and thermal (engine coolant and catalytic converter temperatures) constraints.~{The MPC exploits prediction and optimization over a shorter receding horizon and longer shrinking horizon. The vehicle speed is predicted (or planned in case of eco-driving) based on V2I communications over the shorter receding horizon. Over the longer shrinking horizon, the vehicle speed estimation is based on the data collected from the connected vehicles traveling on the same route as the ego-vehicle.} Simulation results of applying the MPC over real-world urban driving cycles in Ann Arbor, MI are presented to demonstrate the effectiveness and fuel-saving potentials of the proposed iPTM strategy under the uncertainty associated with long-term predictions of the CV’s speed. 

\vspace{-4pt}
\section{Introduction}\label{sec:sec1}\vspace{-8pt}
Hybrid electric vehicles (HEVs) are shown to have major potential in reducing fuel consumption and  %{mitigating the carbon footprint of the transportation sector}
emissions~\cite{axsen2013hybrid}. To fully exploit the fuel economy potential of HEVs, extensive research have been put into design of energy management strategies (EMS)~\textcolor{black}{\cite{malikopoulos2014supervisory,chau2002overview,salmasi2007control,serrao2011comparative}}.~The existing EMS exploit heuristic or optimization-based methods. The heuristic strategies typically rely on rule-based controllers and load leveling logic~\cite{liu2005modeling}, where usually no future driving information is needed.~\textcolor{black}{The optimization-based strategies, on the other hand, can be classified into two major groups: (i) offline approaches, e.g. Dynamic Programming (DP)~\cite{brahma2000optimal,lin2003power} and Pontryagin's Maximum Principle (PMP)~\cite{sciarretta2007control,kim2010optimal,serrao2009ecms}, and (ii) online approaches, e.g., model predictive controllers (MPC)~\cite{borhan2010nonlinear,di2013stochastic,bichi2010stochastic,borhan2011mpc} and Equivalent Consumption Minimization Strategies (ECMS)~\cite{paganelli2001general}.}

\vspace{-4pt} 
%\textcolor{red}
Most of the existing studies on the EMS design are based on the assumption that the engine is operating at normal temperature, i.e., $70-90^oC$~\cite{kim2016thermal}. \textcolor{black}{However,~%{many previous studies have demonstrated that the}
engine thermal management strategies can significantly influence the fuel economy of HEVs~\cite{lohse2013ambient}, especially at cold ambient temperatures.} In cold weather, the engine temperature drops quickly when the vehicle operates in the electric mode (i.e., engine off) and the decreased temperature can degrade the engine performance once it is commanded to turn on.~\textcolor{black}{Furthermore, the cold engine and ambient temperatures affect the aftertreatment system leading to an interaction between fuel economy and emission reduction at low ambient temperatures~\cite{kheir2004emissions}.}~Despite the considerable impact of engine, battery, and aftertreatment thermal management on the HEV fuel economy and efficiency, only a few references~\cite{maamria2017online,zhao2015integrated,wei2019optimal,lescot2010integration,kessels2008towards} have addressed integrated power and thermal management (iPTM) of HEVs.

\vspace{-4pt}
The conventional EMS designed for HEVs at normal operating conditions can be extended to handle iPTM with additional thermal states.~In \cite{wei2019optimal}, influence of the cold-start on the fuel consumption was studied using DP. In \cite{lescot2010integration}, a thermal state was considered reflecting the engine temperature and PMP was applied to a two-state model. DP is computationally demanding for multidimensional optimization problems, and thus is infeasible for real-time implementation. PMP-based approaches, on the other hand, significantly reduce the computational demand as compared to DP, and can be implemented online with real-time adaptation. PMP-based approaches are typically based on simplified models so that the co-state of battery state-of-charge ($SOC$) is constant.~However, extending this approach to include thermal states is difficult as the corresponding co-state is non-constant. Additionally, PMP-based approaches cannot easily handle hard state constraints.~\textcolor{black}{To address these challenges, online approaches have been proposed in previous studies for iPTM. An extended multi-state ECMS is proposed in \cite{maamria2017online}, which can satisfy the emission requirements and provide a sub-optimal solution for fuel consumption. Furthermore, an MPC scheme is developed in~\cite{zhao2015integrated} to reduce catalyst light-off time with minimal impact on the fuel economy.}

\vspace{-4pt}
The main advantage of MPC is in its capacity for handling state and input constraints while approximating optimal feedback control laws. To facilitate MPC, the future vehicle speed prediction is beneficial.%\{prerequisite} 
~For a vehicle driving in a mixed and uncertain traffic environment, any prediction of the future speed is uncertain and this uncertainty can %{degrade robustness}
\textcolor{black}{degrade the performance of MPC}~\cite{fu2011real,cummings2015effect}. Recently, with the advances in vehicle-to-infrastructure (V2I) and vehicle-to-vehicle (V2V) communications, more accurate vehicle speed prediction over a short prediction horizon has become possible~\cite{sun2014dynamic}. Long-range vehicle speed prediction, however, is still difficult and uncertain. For iPTM of HEVs,~%{compared with power systems, thermal systems usually have relatively slow dynamics.} 
the need to handle slow responding thermal dynamics calls for a long prediction horizon to achieve good performance. Since a long-term accurate vehicle speed preview is often not available, MPC for iPTM of HEVs is typically implemented with a short prediction horizon, which limits its energy saving.

\vspace{-4pt}
\textcolor{black}{In our previous study~\cite{gong2019integrated}, we developed a DP-based iPTM strategy with consideration of engine coolant and cabin heating demands for HEVs.~Assuming an accurate prediction of the future vehicle speed, the DP-based iPTM illustrated the benefits of leveraging vehicle speed preview, when coordinating power and thermal states to optimize fuel economy subject to thermal demands.~In this paper, we develop a new MPC-based approach for real-time iPTM, with consideration of engine coolant and aftertreatment systems for HEVs. In order to enhance the robustness of the MPC-based iPTM strategy against the uncertainties associated with vehicle speed forecasts, a multi-range speed prediction and planning scheme is proposed in this paper}.~%{The proposed MPC-based iPTM approach leverages both short- and long-range speed previews with different accuracies.} 
In this approach, both short- and long-range speed previews with different accuracies are leveraged. The short-range speed prediction is obtained via V2I/V2V data. The long-range speed prediction is estimated based on analysis of historical traffic data collected from connected vehicles (CVs) driving along the same corridor as the ego-vehicle. Moreover, the proposed MPC strategy is implemented for a winter driving condition when the cabin heating demand is high. The objective of iPTM is to minimize the fuel consumption, while enforcing the power and thermal (engine coolant and catalytic converter temperatures) constraints, as well as satisfying the charge sustaining ($SOC$) requirement. {Furthermore, simplified prediction models are required to predict the vehicle power and thermal dynamics for real-time optimization. To this end, control-oriented models of the engine and battery $SOC$ are adopted from our previous work~\cite{gong2019integrated}, and a new control-oriented catalyst temperature model is developed and experimentally validated in this paper.%}

\vspace{-4pt}
{The main contributions of this paper are: (i) development and experimental validation of a control-oriented catalyst temperature model of a power-split HEV aftertreatment system, (ii) development of an MPC-based iPTM strategy to leverage a multi-range speed preview enabled by connectivity technology, and (iii) investigating the sensitivity of the MPC-based iPTM to long-term vehicle speed forecast errors. While the uncertainty in the vehicle speed prediction affects the iPTM results unfavorably, it is shown that an approximate knowledge of future vehicle speed that captures the main traffic events (e.g., stops and acceleration events) is sufficient to achieve fuel economy improvement of connected HEVs.}

\section{HEV Power and Thermal Systems Modeling}\label{sec:sec2}  \vspace{-8pt}
The schematic of a power-split HEV with power and thermal loops is shown in Figure~\ref{fig:Thermal_System_TWC}. The traction power demand ($P_{trac}$) is provided by blending power from the internal combustion engine ($P_{eng}$) and electric battery ($P_{bat}$) %{at different ratios determined by}
through the power-split device (PSD). In the thermal loop, the coolant is used to regulate the engine temperature ($T_{eng}$).{~It is assumed that any combustion energy not converted to mechanical work is either transferred to the engine coolant, transferred to the catalyst, or leaves the system via the exhaust.}~The stored thermal energy in the coolant is released to the ambient (via the radiator and air convection), and to the cabin compartment (via the heater cores), depending on the cabin heating system~%{(i.e., air conditioning)}
demands and coolant temperature ($T_{cl}$). The thermal dynamics of the three-way catalytic converter (TWC) temperature ($T_{cat}$) is coupled with the engine power and thermal responses, and is dominated by the exhaust gas temperature and flow rate. To design an optimization-based iPTM strategy, simplified models capturing the HEV power and thermal dynamics associated with the battery $SOC$, $T_{cl}$, and $T_{cat}$ are required.~These models are introduced in the following subsections. \vspace{-4pt}
\begin{figure}[h!]
\begin{center}
\includegraphics[angle=0,width= \columnwidth]{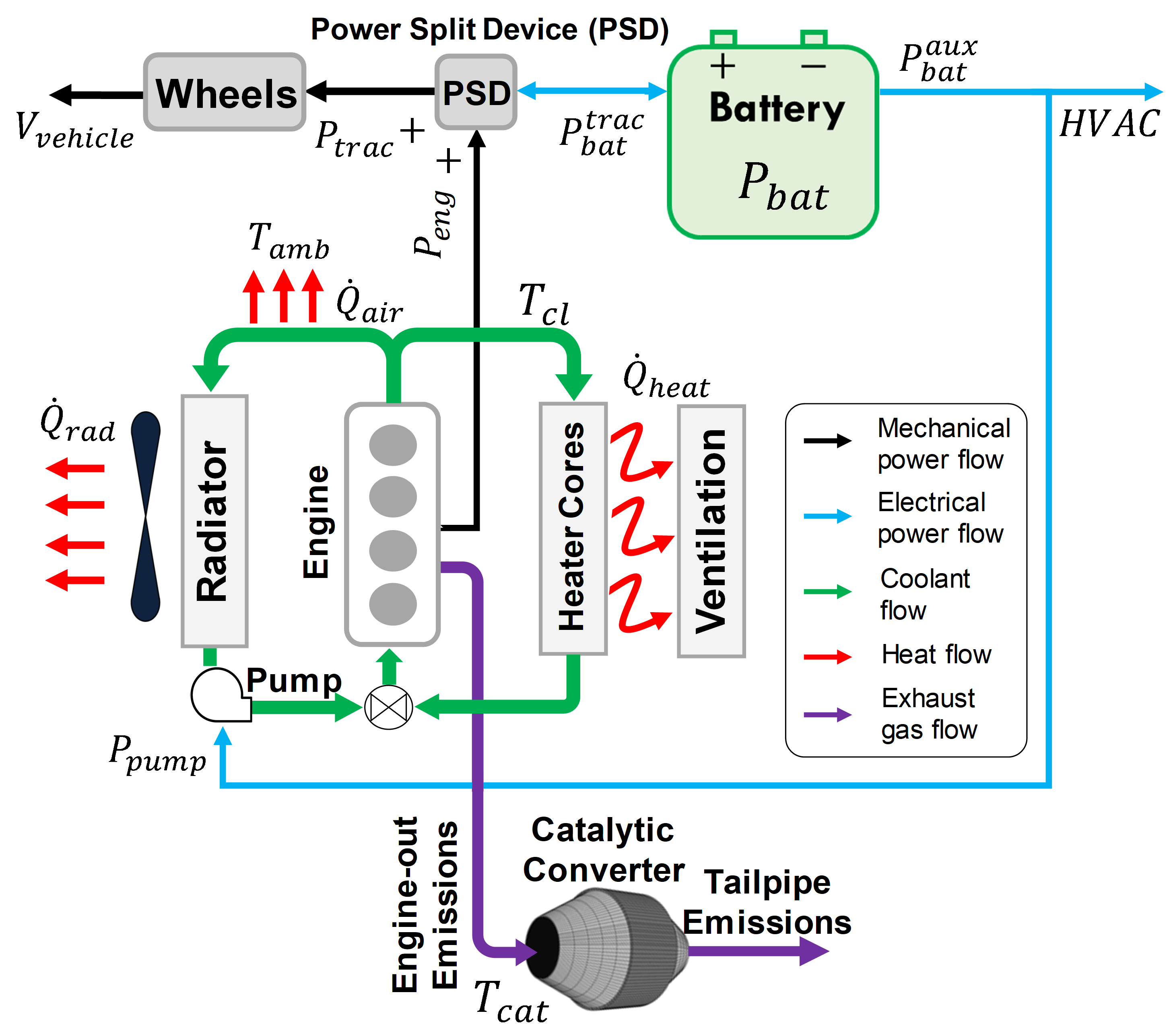} \vspace{-0.75cm}
\textcolor{Blue}{\caption{\label{fig:Thermal_System_TWC} HEV power and thermal systems coupled with exhaust aftertreatment system.}}\vspace{-4pt}
\end{center}
\end{figure}

\subsection{Battery Power-Balance Model}\label{subsec:sec1_1}\vspace{-8pt}
The battery contributes electrical power to satisfy the traction power demand for driving, i.e.,
\vspace{-0.15cm}
\begin{gather}\label{eq:Power_demand}
P_{bat}^{trac}(t)=P_{trac}(t)-P_{eng}(t),
\end{gather}
where %{$P_{trac}=\omega_{e}\tau_{e}$ is the total traction power demand, and }
$P_{eng}$ is the engine mechanical output power determined by engine speed ($\omega_{e}$) and torque ($\tau_{e}$), which follow the optimal operating points line on which engine brake specific fuel consumption (BSFC) is minimized. In addition to the traction power, the battery provides power to other auxiliary loads ($P_{bat}^{aux}$) across the vehicle, e.g., HVAC system, engine coolant pump, etc. An equivalent circuit model is used to capture the $SOC$ dynamics:
%
%\small
\vspace{-0.15cm}
\begin{gather}\label{eq:SOC_simple_model}
\dot{SOC}(t)=\frac{U_{oc}(t)-{\sqrt {U_{oc}^2(t)-4R_{int}(t)P_{bat}(t)} }}{{2{R_{int}(t){C_{bat}}}}},
\end{gather}
where $P_{bat}=P_{bat}^{trac}+P_{bat}^{aux}$, and $C_{bat}$, $R_{int}$ and $U_{oc}$ are the battery power, capacity, internal resistance, and open-circuit voltage, respectively. The battery $SOC$ model in Eq.~(\ref{eq:SOC_simple_model}) has been validated against the data collected from an HEV in our previous work, see~\cite{gong2019integrated} for more details and validation results. 

\vspace{-4pt}
\subsection{Engine Coolant Thermal Model}\label{subsec:sec1_2}\vspace{-8pt}
We developed and experimentally validated a control-oriented model of the coolant temperature ($T_{cl}$) in our previous work~\cite{Amini_CCTA19,gong2019integrated}. The structure of the model is shown
below: %\textcolor{red}{The validated dynamic model is used in this work and the structure of the model is shown below:}
\vspace{-0.35cm}
\begin{gather}\label{eq:thermal_temp}
\dot{T}_{cl}(t)=\frac{1}{M_{eng}C_{eng}}({\dot{Q}}_{fuel}-P_{eng}-{\dot{Q}}_{exh}-{\dot{Q}}_{air}-{\dot{Q}}_{heat}),
\end{gather}
where $M_{eng}$ and $C_{eng}$ are the equivalent thermal mass and capacity of the engine cooling system, respectively.~{Additionally, ${\dot{Q}}_{fuel}$, ${\dot{Q}}_{exh}$, ${\dot{Q}}_{air}$ and ${\dot{Q}}_{heat}$ are the heat rate released in the combustion process, rejected in the exhaust, rejected by air convection and exchanged for cabin heating, respectively, see Figure.~\ref{fig:Thermal_System_TWC}. In Eq.~(\ref{eq:thermal_temp}), ${\dot{Q}}_{fuel}$ is calculated as a function of the fuel consumption rate ($\dot{m}_{fuel}$) and lower heating value ($LHV$) of gasoline:}
\vspace{-0.15cm}
\begin{gather}\label{eq:heat_fuel}
{\dot{Q}}_{fuel}=LHV\cdot \dot{m}_{fuel}(\omega_{e},\tau_{e},T_{cl})%\\
\end{gather}
where $\dot{m}_{fuel}$ is modelled as a function of engine speed ($\omega_{e}$), engine torque ($\tau_{e}$) and engine coolant temperature:\vspace{-0.15cm}
\begin{gather}
\label{eq:fuel_rate}
\dot{m}_{fuel}(\omega_{e},\tau_{e},T_{cl})=\alpha(T_{cl})\cdot f_{fuel}(\omega_{e},\tau_{e})
\end{gather}
in which $f_{fuel}(\omega_{e},\tau_{e})$ is the nominal fuel consumption rate calculated according to the BSFC map and $\alpha(T_{cl})$ is a correction multiplier introduced to reflect the impact of $T_{cl}$. The function $\alpha(T_{cl})$ can be found in Autonomie~\cite{kim2014thermal} software's thermal HEV model developed by Argonne National Laboratory (ANL), see~\cite{gong2019integrated}. Figure~\ref{fig:Tcl_Model_ID_Vrf} shows the validation results of $T_{cl}$ model (Eq.~(\ref{eq:thermal_temp})) using experimental
data collected from the test HEV over highway (Figure~\ref{fig:Tcl_Model_ID_Vrf}-($a_{1,2}$)) and  city (Figure~\ref{fig:Tcl_Model_ID_Vrf}-($b_{1,2}$)) driving routes in Ann Arbor, MI.\vspace{-10pt} 
\begin{figure}[h!]
\begin{center}
\includegraphics[angle=0,width=0.95\columnwidth]{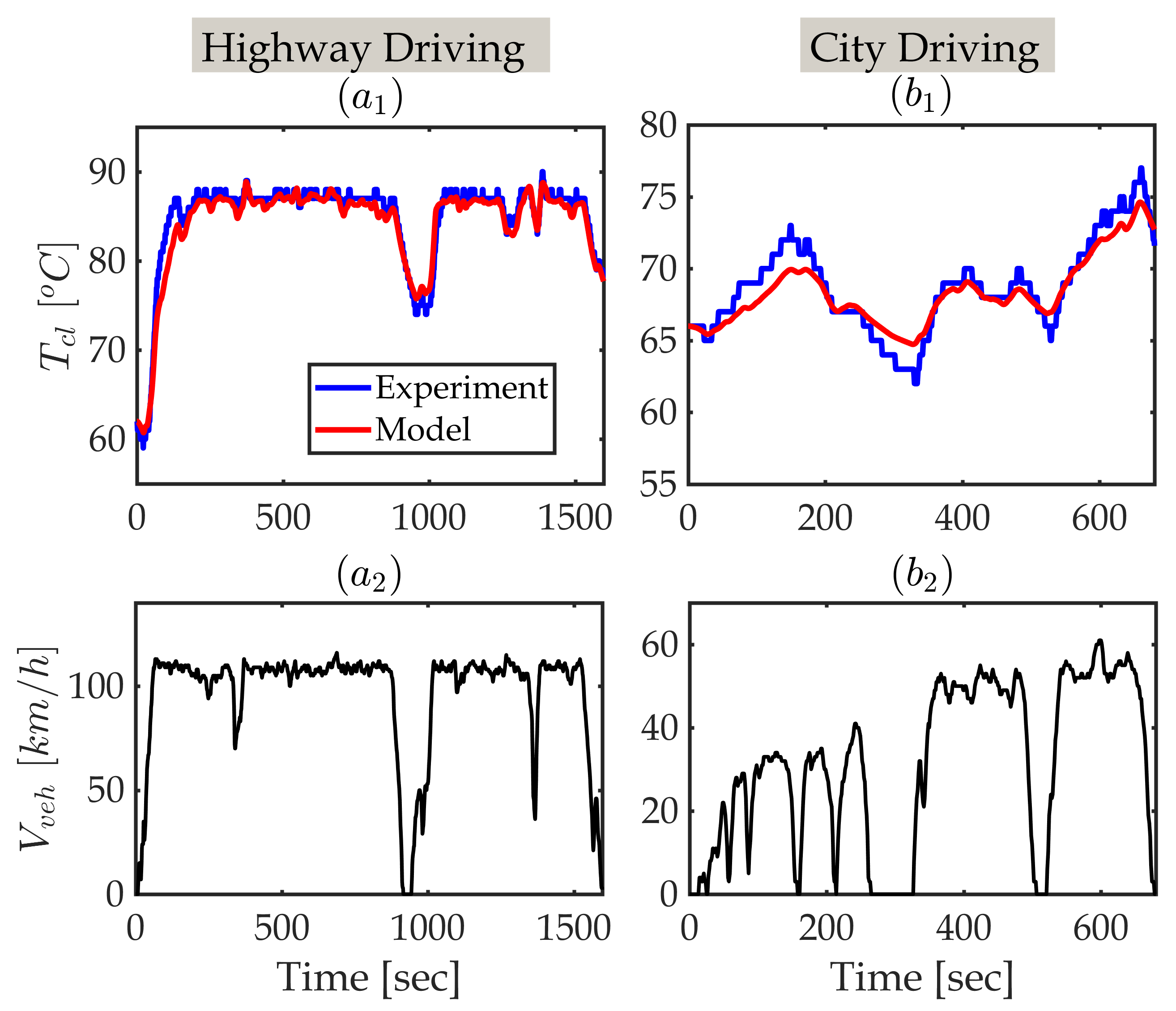} \vspace{-8pt}
\textcolor{Blue}{\caption{\label{fig:Tcl_Model_ID_Vrf} The results of coolant temperature ({$T_{cl}$}) model validation.}}\vspace{-6pt}
\end{center}
\end{figure}

\subsection{Catalytic Converter Thermal Model}\label{subsec:sec1_3}\vspace{-8pt}
\begin{figure*}%[h!]
\begin{center}
\includegraphics[angle=0,width=1.85 \columnwidth]{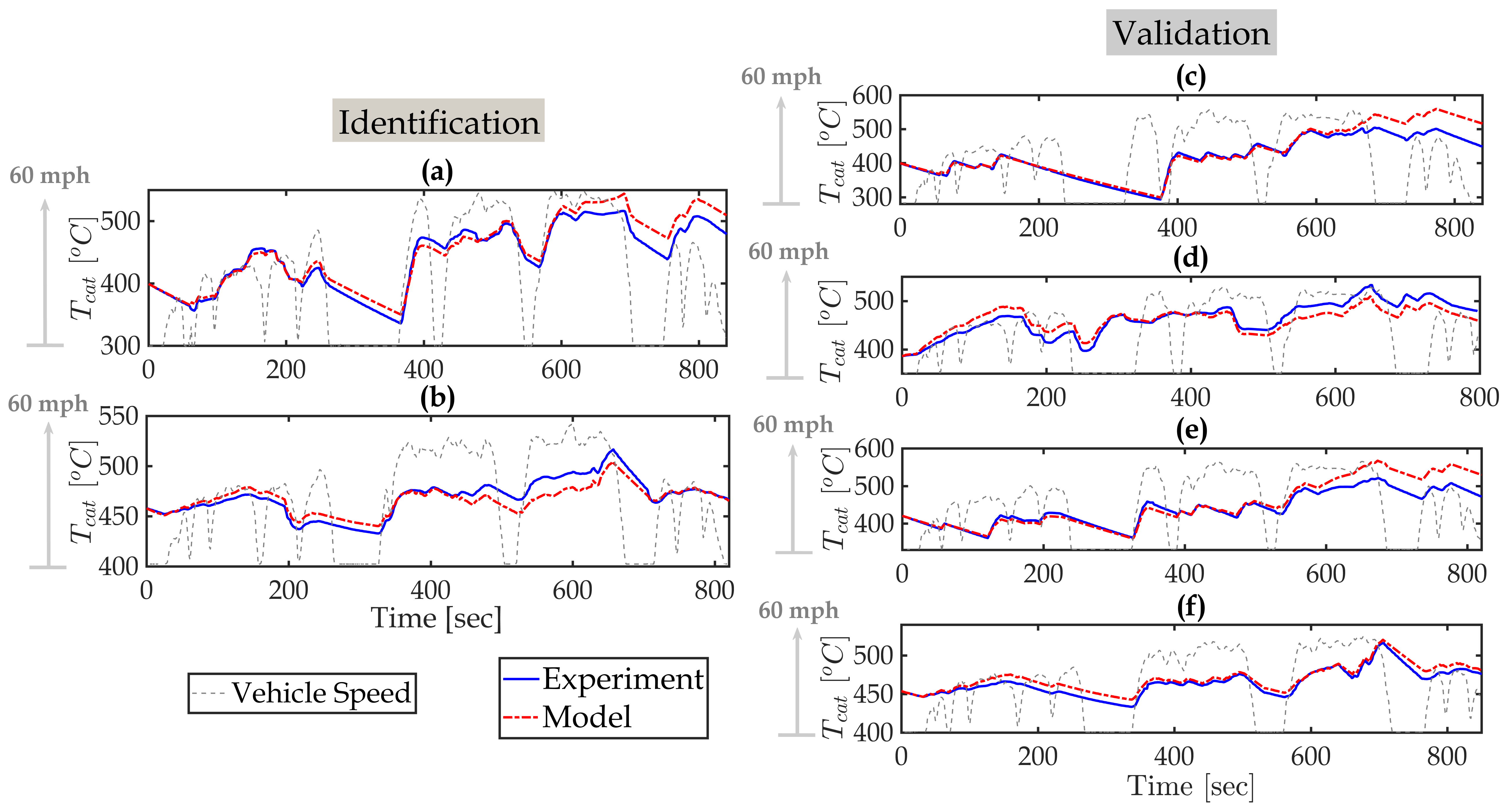} \vspace{-0.5cm}
\textcolor{Blue}{\caption{\label{fig:T_cat_id_verf} Identification and validation results of the control-oriented switching model of the TWC temperature: (\textbf{a,b}) identification, (\textbf{c,d,e,f}) validation. The experimental data are collected from a power-split HEV driven in Ann Arbor, MI.}}\vspace{-6pt}
\end{center}
\end{figure*}
A control-oriented model for $T_{cat}$ is developed and experimentally validated. The data used for identification and validation of the proposed model are collected from a test power-split HEV, which is driven in Ann Arbor, MI.~The TWC has highly nonlinear dynamics and its response varies depending on the operation modes of the combustion engine, i.e., on, off, idling. Such thermal behavior is difficult to be captured using a single control-oriented model. To this end, here, we propose a switching model to predict $T_{cat}$ during engine on and off modes. {Note that we showed in our previous work~\cite{gong2019integrated} that the optimal operation of the power and thermal systems requires no engine idling, when there is no relatively long stop along the driving cycle. For the specific urban driving cycles considered in this paper, we do not model the TWC thermal response during engine idling, assuming that the engine will not idle if controlled through an optimization scheme. This assumption holds for the scenarios presented in the following sections.}

\vspace{-4pt}
The proposed switching $T_{cat}$ has the following form:\vspace{-4pt}
\begin{gather}\label{eq:T_cat_on}
%\small
\dot{T}_{cat}(t)=
\begin{cases}
%\scriptstyle
(\alpha_1+\alpha_2V_{veh})(T_{cat}-T_{amb})+\alpha_3\omega_e+\alpha_4{\omega_e}^2+\\\alpha_5\tau_e+\alpha_6{\tau_e}^2+\alpha_7{\omega_e}^2\tau_e+\alpha_8,~~~~\textbf{if engine On}\\
~\vspace{-0.25cm}\\
\beta_1(T_{cat}-T_{amb})+\beta_2,~~~~~~~~~~~~~~~~\textbf{if engine Off} 
\end{cases}%\nonumber
\end{gather}
where $V_{veh}$ and $T_{amb}$ are the vehicle speed and the ambient temperature, respectively. When the engine is on, $\dot{T}_{cat}$ is determined by heat convection, engine speed ($\omega_e$) and torque ($\tau_e$). The heat convection is proportional to the temperature difference between $T_{cat}$ and $T_{amb}$ with a vehicle speed dependent coefficient. Additionally, the linear and quadratic terms of engine torque and speed are used to represent the impact of engine operating condition. When the engine is off, the thermal behavior is dominated by the heat convection with ambient air. {Note that the analysis of the experimental data revealed that $\beta_1$ in (\ref{eq:T_cat_on}) can be identified as a constant with negligible dependency on the vehicle speed. As an example, Figure \ref{fig:T_cat_id_verf}-(c) shows that from $t=150$ to $380~sec$ the engine is off while the vehicle speed varies in a large range. Despite these variations in the vehicle speed, Figure \ref{fig:T_cat_id_verf}-(c) confirms that the proposed $T_{cat}$ model with constant $\beta_1$ matches the experimental data well.}
\vspace{-14pt}
\begin{itemize}
    \item \textcolor{black}{\textbf{Remark 1}: the proposed $T_{cat}$ model in Eq.~(\ref{eq:T_cat_on}) contains both engine speed and torque terms as the experimental data used for identification of the model parameter are collected from a test vehicle in which the engine does not necessarily follow the optimal operating line (OOL). Although, the following controller implementation and simulations, we assume the engine follows the OOL, Eq.~(\ref{eq:T_cat_on}) may support future off-OOL studies.}
\end{itemize}
    
\vspace{-10pt}
The parameters of the $T_{cat}$ model in Eq. (\ref{eq:T_cat_on}). i.e., $\alpha_1,\cdots,\alpha_8,\beta_1,\beta_2$, are identified using constrained least-square approach. These parameters are listed in the Appendix. The identification and validation results are presented in Figure~\ref{fig:T_cat_id_verf}, where subplots ($a,b$) show the data used for parameter identification, and ($c,d,e,f$) present the model validation results. \textcolor{black}{The data in Figure~\ref{fig:T_cat_id_verf} is collected from the test vehicle while repeatedly driving the same route in Ann Arbor, MI. Each journey is subject to different traffic patterns, ambient temperatures and initial powertrain states.  Datasets (a) and (b) were randomly selected for model calibration, while (c)-(f) were used as validation cases.} Table~\ref{tab:Different_timescale} summarizes the proposed model's accuracy in terms of the average absolute error and standard deviation. \textcolor{black}{Given that the TWC temperature predictions are made at $t=0~sec$ with known initial $T_{cl,init}$, the validation results and Table~\ref{tab:Different_timescale} shows that the proposed model delivers acceptable prediction accuracy, with normalized average absolute error of 3.6\% ($c$), 3.1\% ($d$), 3.9\% ($e$) and 1.2\% ($f$). Overall, our investigation showed that the impact of up to 4.0\% normalized average absolute error on the controller performance is negligible, this has been considered as the “acceptable” accuracy.}

\begin{table}[!h]
\centering
\vspace{-10pt}
\textcolor{Blue}{\caption{The average absolute error and standard deviation of the control-oriented $T_{cat}$ model in predicting the actual catalyst temperature based on the validation results shown in Figure~\ref{fig:T_cat_id_verf}-(c,d,e,f). \label{tab:Different_timescale}}}\vspace{0.25cm}
\begin{tabular}{ccccc}
\hline
subplot in Figure~\ref{fig:T_cat_id_verf} & (\textbf{c}) & (\textbf{d}) & (\textbf{e}) & (\textbf{f}) \\
\hline
\hline
avg. absolute error [$^oC$] & 14.9 & 14.1 & 17.1 & 5.6\\
standard deviation [$^oC$] & 21.8 & 15.8& 22.1 & 3.3\\
\hline
\end{tabular}\vspace{-10pt}
\end{table}

\vspace{-8pt}
\section{Multi-Range Speed Prediction for Connected HEVs}\vspace{-8pt}
\begin{figure*}[h!]
\begin{center}
\includegraphics[angle=0,width=1.15\columnwidth]{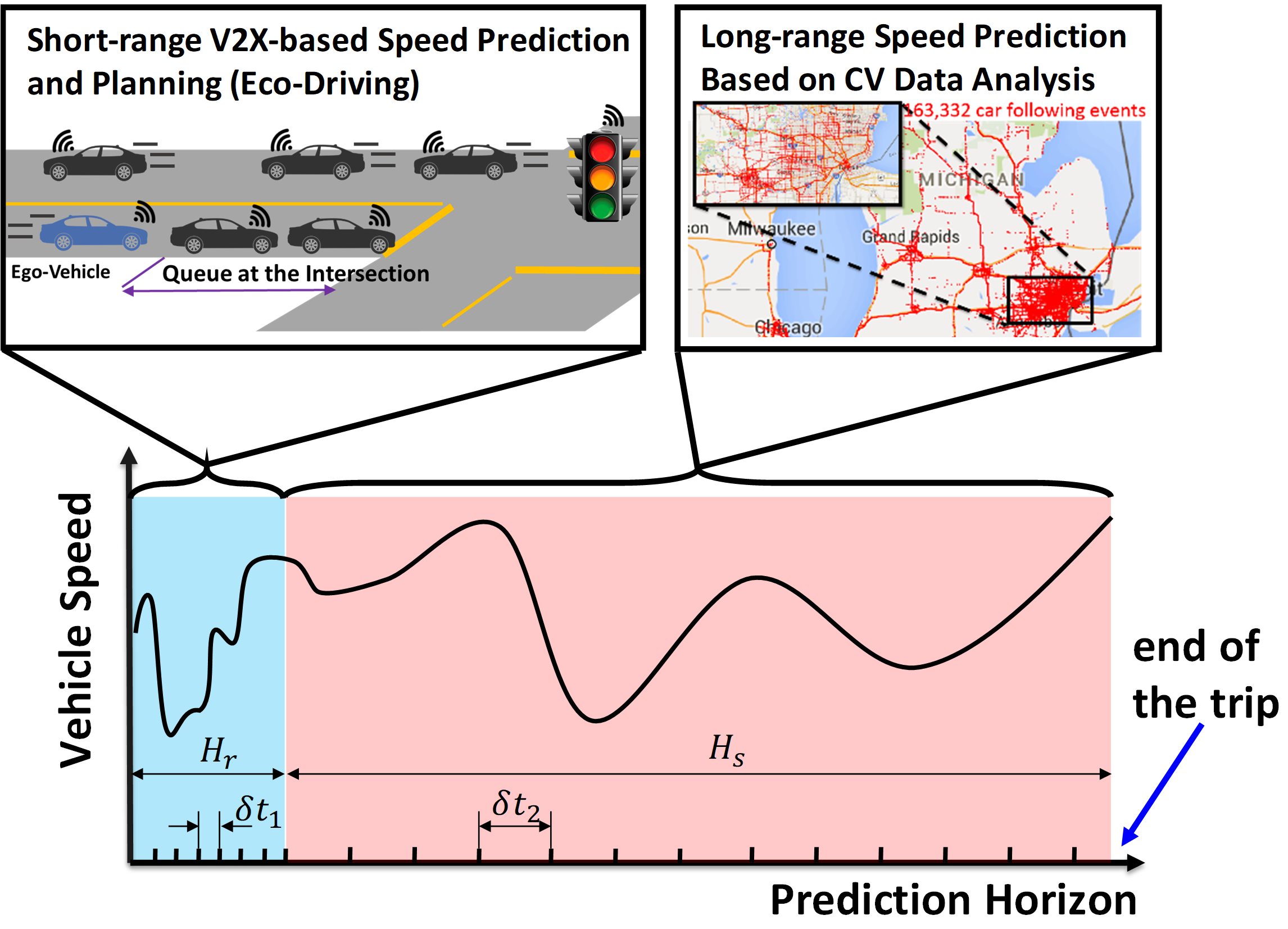} \vspace{-0.5cm}
\textcolor{Blue}{\caption{\label{fig:MH_concept} The schematic of the multi-horizon ego-vehicle speed preview with short and long-range prediction horizons. The short-range speed prediction is based on V2X communication, while the long-term prediction is realized through analyzing the traffic data from connected vehicles (CV) traveling along the same route as the eco-vehicle. $H_r$ and $H_s$ denote receding and shrinking prediction horizons, respectively.}}
\end{center}\vspace{-4pt}
\end{figure*}

As discussed in the introduction section, the slow dynamics of the thermal systems call for a relatively long prediction horizon to achieve the best performance via the optimization-based strategies. %{Connectivity has made short-range prediction of the vehicle speed more accessible.}
Long-range vehicle speed predictions, however, are still error-prone.~Nevertheless in~\cite{AminiCDC18,AminiACC19,Amini_TCST_2019}, it was shown that even an approximate knowledge of major traffic events and trends can be leveraged for slow responding thermal systems, e.g., battery thermal management and cabin air conditioning, for improving the energy efficiency. \vspace{-4pt}

In order to maximize the use of connectivity-enabled data from different resources, a multi-range speed prediction scheme is implemented for iPTM of connected HEVs. The speed prediction scheme %{consists of two parts with different prediction accuracies:\vspace{-14pt}}
is performed over both short and long horizons, as illustrated in Figure~\ref{fig:MH_concept}:\vspace{-14pt}
\begin{itemize}
    \item \textbf{Short-range speed prediction}: {The short range (e.g. $5-20~sec$ ahead) is based on the V2I/V2V (V2X) communications available to the connected HEVs and is assumed to be highly accurate. It specifies vehicle speed values at %{a fast update rate }
    a high resolution of $\delta t_1~sec$}.
    \vspace{-6pt}
    \item \textbf{Long-range speed prediction}: {It is assumed that the long-range vehicle speed forecast beyond the short-range prediction window and till the end of the trip is available. This forecast is less accurate and specifies vehicle speed values at a %{slower update rate }
    low resolution of $\delta t_2~sec$.}
\end{itemize}\vspace{-14pt}
The proposed multi-range speed prediction scheme is further discussed in the following subsections. Firstly, the short-range V2X-based speed prediction algorithm is briefly introduced. This algorithm is based on our previous work~\cite{yang2019eco,AminiACC19}, in which eco-trajectory speed prediction and planning were developed for eco-driving. Secondly, a data %{clustering}
classification algorithm is applied to the data collected from all the connected vehicles (CVs) driving along the same route as the ego-vehicle. This data classification allows for estimation of the average speed of the vehicle over a much longer horizon, as compared to the V2X-based speed predictions. It is assumed that the end of the trip can be predicted over the long prediction horizon. Thus, as the vehicle proceeds along the planned route, the time horizon---over which vehicle speed forecast is generated---is shrinking. 

\vspace{-4pt}
\subsection{Short-Range Speed Prediction {via} Eco-trajectory Planning}\vspace{-8pt}
\begin{figure*}%[b]
\begin{center}
\includegraphics[angle=0,width= 1.55\columnwidth]{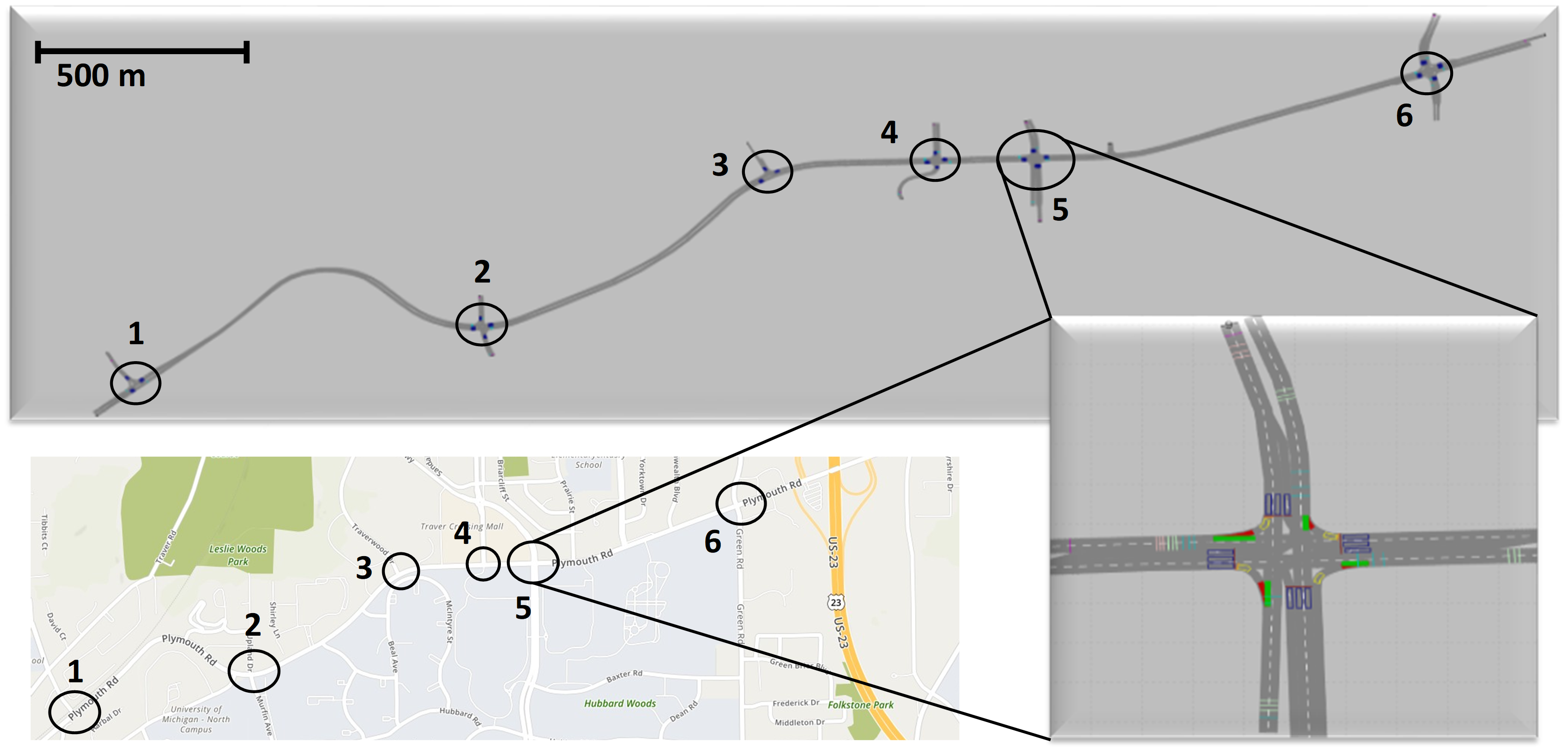} \vspace{-0.4cm}
\textcolor{Blue}{\caption{\label{fig:PlymouthRd_map} Plymouth corridor in Ann Arbor for traffic modeling and simulation.}}
\end{center}
\end{figure*}
Over the short-range vehicle speed prediction window, it is assumed that the vehicle speed can be predicted accurately based on the V2X data. Moreover, in case of eco-driving, an eco-trajectory planning algorithm can be used to optimize the vehicle speed trajectory which the ego-vehicle will follow. In this eco-driving case, the vehicle is assumed to be able to accurately follow the planned speed trajectory. The eco-trajectory planning approach used in this paper, which is based on our previous works~\cite{yang2019eco,AminiACC19}, accounts for queuing dynamics along congested corridors. In this framework, the eco-vehicle receives traffic signal and queue length information via V2I communications and generates a speed profile with the objective of minimizing energy consumption. The queue length is predicted based on the trajectories of connected vehicles inferred from Basic Safety Messages (BSMs) and from loop-detectors installed on the infrastructure side. The queuing process is modeled based on the shockwave profile model (SPM)~\cite{wu2011shockwave} to provide a green window (i.e., the time interval during which an eco-driving vehicle can pass through a given intersection) for eco-driving trajectory planning. Our algorithm is able to predict the queuing dynamics and estimate the green window before the eco-driving vehicle's arrival at the intersection. The thereby planned vehicle speed trajectory for the ego-vehicle can be used as the vehicle speed forecast for iPTM.
 
\vspace{-4pt}
{The eco-trajectory planning algorithm is applied in microscopic traffic simulation software VISSIM~\cite{ptv2016ptv} to a six-intersection corridor on Plymouth Rd. in Ann Arbor, MI, as shown in Figure~\ref{fig:PlymouthRd_map} with the black circles denoting the location of the intersections. The stretch of the considered road segment is about 2.2 miles long and has two lanes in each direction. This stretch is one of the busiest local commuting routes, connecting US23 to the North Campus of the University of Michigan and downtown Ann Arbor. To calibrate the VISSIM simulation model and represent a congested traffic condition, real-world data have been collected during PM rush hour (4:00-5:00PM), including traffic volume, turning ratio, and traffic signal timing at each intersection. All the traffic signals and vehicles are programmed to broadcast signal status in real-time. These data are sent to the queuing profile algorithm for green window prediction, which is then used in the eco-trajectory planning algorithm, ~see~\cite{barth2011dynamic,yang2019eco,AminiACC19}.

\vspace{-8pt}
\subsection{Long-Range Speed Prediction {via} Traffic data mining }\vspace{-8pt}
For long-term speed prediction, we are specifically interested in the forecast of the major traffic events based on the current states of the vehicle, infrastructure, and traffic data. These major traffic events may include the average vehicle cruise speed between the intersections, relatively long stops, and significant acceleration events. Such~long-range prediction could be realized by analysing extensive GPS-based position and velocity measurements from the vehicles travelling on the same route and estimate the average traffic flow speed~\cite{herrera2010evaluation,sun2015integrating}. \textcolor{black}{However, this approach may not accurately capture traffic patterns in city driving with congestion and multiple intersections.}

\vspace{-4pt}
In order to estimate the long-term speed trajectories, a data mining algorithm is used in this paper to analyze and classify the traffic data from the connected vehicles. For a city driving cycle, the traffic signals on arterial corridors greatly influence the traffic flow with the stop-and-go feature.~%, which has a significant impact on the fuel economy. On the other hand, the traffic signals regulate different traffic flows to follow certain patterns, defined by the signal timing plan. 
Assuming that the traffic signal information is known a \textit{priori}, the vehicle speed trajectories can be classified based on the signal timing plans. To this end, the VISSIM model for the same six-intersection corridor shown in Figure~\ref{fig:PlymouthRd_map} is run~%{for four hours} 
and the trajectories of mixed traffic (1531 vehicles in total, 50 of which implement eco-driving) are collected over the course of four hours. Note that, while the majority of the vehicles do not implement eco-driving, all of them are connected and they can communicate their speed and position data. Next, a rule-based data classification algorithm is applied to the data collected from all the vehicles traveling on the same route as the ego-vehicle. These CVs speed data are categorized into 10 bins based on their arrival time at the first intersection shown in Figure~\ref{fig:PlymouthRd_map}. One signal cycle of 100 $sec$, which begins with the signal turning red, is divided equally into 10 intervals and each interval consists of 10 $sec$ corresponding to one bin.~For example, if a vehicle arrives at the intersection 45 seconds after the signal turns to red, the vehicle is classified into bin \#5.
\begin{figure*}%[h!]
\begin{center}
\includegraphics[angle=0,width=1.52\columnwidth]{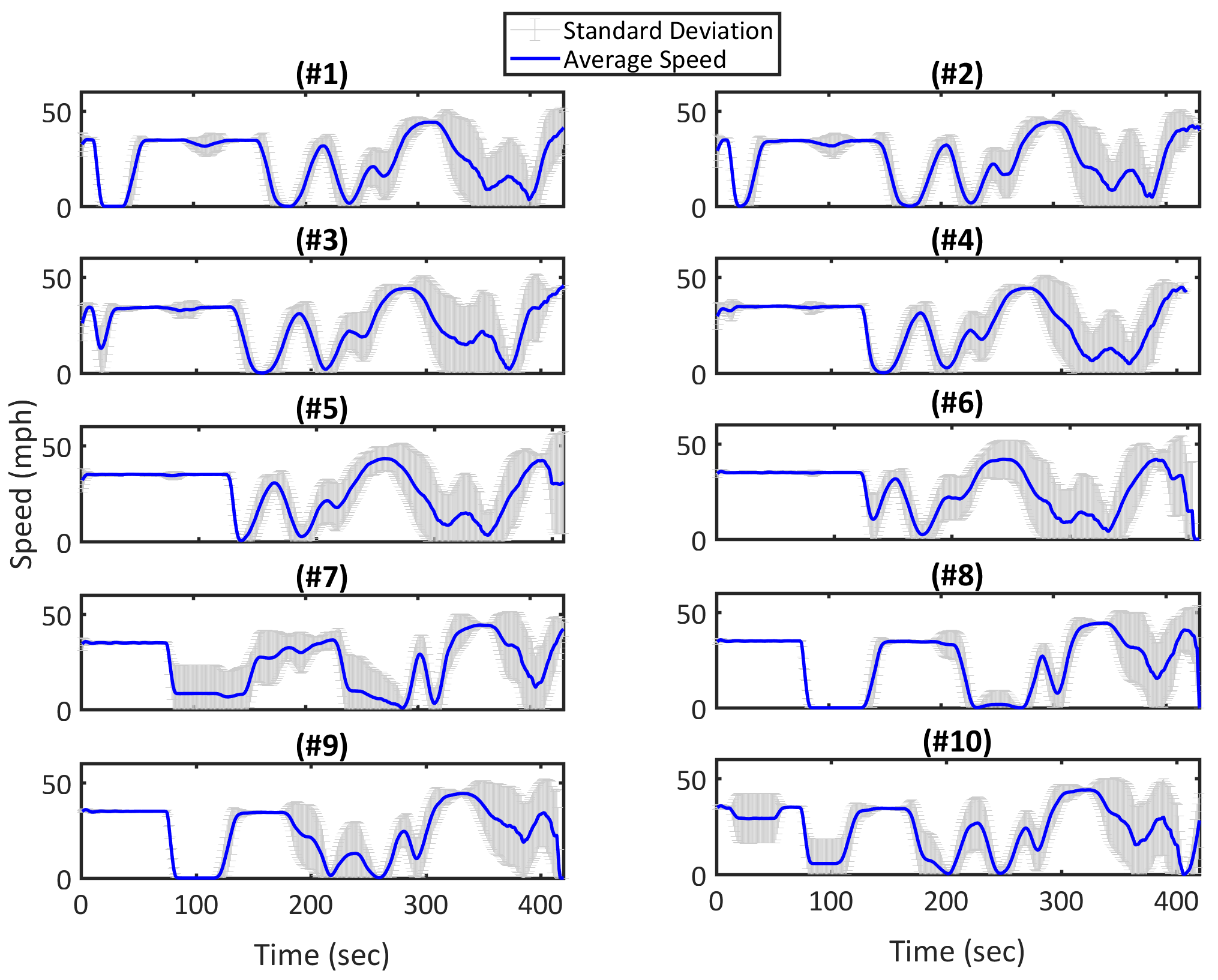} \vspace{-0.5cm}
\textcolor{Blue}{\caption{\label{fig:FinalFig_1to10_withAggregated} Average and standard deviation of the “classified” speed profiles in 10 bins.}}\vspace{-4pt}
\end{center}
\end{figure*}

\vspace{-14pt}
\begin{itemize}
    \item \textcolor{black}{\textbf{Remark 2}: according to Figure~\ref{fig:PlymouthRd_map}, the first intersection is close to the starting location of the trip. It is assumed that the first intersection is within the range of the V2I communication and the speed profile can be predicted accordingly over the short-range before arriving at the first intersection. By knowing the signal cycle at the first intersection and the predicted vehicle speed, the arrival time at the first intersection and the associated bin number of the vehicle can be predicted as well. Thus, the long-term vehicle speed can be estimated before arriving at the first intersection.}
\end{itemize}

\vspace{-8pt}
The average and standard deviation of the vehicle speed profiles clustered into these 10 bins are shown in Figures~\ref{fig:FinalFig_1to10_withAggregated}-(\#1 to \#10). Figure~\ref{fig:FinalFig_1to10_withAggregated}} confirms that the applied data classification strategy can capture the major trends in the traffic flow. Since the vehicle classification is done only based on the arrival time at the first intersection, the speed variations increase spatially. It is noted that the speed variations are different in different bins. {This means that depending on the arrival time at the first intersection, the long-range speed prediction of the ego-vehicle speed could be associated with small (e.g., bin \#8) or large (e.g., bin \#7) uncertainties. Since the long-term predictions are based on the arrival time at the first intersection, the uncertainty increases in all bins as the ego-vehicle approaches the following intersections. 
\vspace{-14pt}
\begin{itemize}
    \item \textcolor{black}{\textbf{Remark 3}: the traffic signals along the Plymouth corridor are assumed to be coordinated, meaning all the signals have a common cycle length but different offsets. The speed variations (e.g., stops and stop time) change gradually across bins. For example, the vehicle stops at the first intersection in Bin \#1 for around $20~sec$ between $t=20-40~sec$, but it stops for a shorter time in Bin \#2 (only stops for a few seconds). Then the vehicle does not need to stop in Bin \#3, and it only slows down. Starting from Bin \#4, the vehicle does not need to slow down anymore at the first intersection. The same trend happens at the following intersections. Due to this coordination, the number of signal patterns that a vehicle can experience is limited;~%{even if it arrives at different portions of a signal cycle}
    this is why speed profiles in certain bins (Figures~\ref{fig:FinalFig_1to10_withAggregated}) are similar.}\vspace{-6pt}
    \item \textcolor{black}{\textbf{Remark 4:} at the beginning of the trip, the prediction of the end time may be different from the actual trip end time. However, as the vehicle approaches the end of the actual trip, the vehicle speed can be predicted more accurately, which helps to estimate the actual trip end time. Since the MPC knows the actual end time of the trip, it is able to enforce the $SOC$ charge sustaining constraint.}
\end{itemize}

\vspace{-4pt}
\section{{i}PTM {of} HEV {via} Model Predictive Control}\vspace{-4pt}
In order to leverage the multi-range speed prediction strategy developed in the previous section, an MPC-based approach is designed in this section. The objective of the iPTM strategy is to minimize the total trip fuel consumption while enforcing the power and thermal constraints (i.e., $SOC$, $T_{cl}$, $T_{cat}$ limits) and meeting the traction and cabin heating demands. {The multi-range speed preview calls for a multi-horizon optimization framework that can use the vehicle speed prediction with different resolutions over different prediction horizons. To this end, a multi-horizon MPC is formulated with the same multi-range prediction horizon shown in Figure~\ref{fig:MH_concept}. 

\vspace{-4pt}
We consider an %{multi-horizon}
MPC with an economic cost function %{($\dot{m}_{fuel}$) }
defined over multiple horizons with battery power, $P_{bat}=P_{bat}^{trac}+P_{bat}^{aux}$, being the optimization variable: 
\vspace{-6pt}
\begin{equation}\label{eq:variable_timescale_MPC_formulation}
\begin{split}
\text{min}~\ell = \text{min}
\{\sum_{i=t}^{t+H_r-1}{\dot{m}}_{fuel}\big(P_{eng}(i),T_{cl}(i)\big)\delta t_1~~\\
+\sum_{j=t+H_r}^{t_{end}}{\dot{m}}_{fuel}\big(P_{eng}(j),T_{cl}(j)\big)\delta t_2\}%\\
%\int_t^{t+N}{\dot{m}}_{fuel}(P_{eng}(\tau),T_{cl}(\tau))d{\tau}\\+\int_{t+N}^{t_{end}}{\dot{m}}_{fuel}(P_{eng}(\tau),T_{cl}(\tau))d{\tau}
\end{split}
\end{equation}
subject to:\vspace{-0.45cm}
\begin{gather}
0.4 \le SOC(k) \le 0.8,\\
40^oC\le T_{cl}(k) \le 90^oC,\\
%\end{gather}
%\begin{gather}
T_{cat}^{min}\le T_{cat}(k),\label{eq:Tcat_constraint}\\
P_{eng}(k)=P_{trac}(k)-P_{bat}(k),\label{eq:Power_constraint}\\
SOC_{init}=SOC_{end},\label{eq:SOC_constraint}%\\
\end{gather}
where $k=\{i,j\}$, $H_r$ is the short moving (receding) horizon (see Figure~\ref{fig:MH_concept}), $t$ indicates the current time, and $t_{end}$ is the final time of the trip.~% The parameters in the state constraints are: $SOC_{min}=0.4$, $SOC_{max}=0.8$, $T_{cl,min}=40^oC$ and $T_{cl,max}=90^oC$.
{In order to reduce the computation time, the prediction horizon is sampled with different resolutions, i.e., $\delta t_1=1~sec$ and $\delta t_2=10~sec$. \textcolor{black}{Similarly, the predictive models of $SOC$ (Eq.~(\ref{eq:SOC_simple_model})), $T_{cl}$ (Eq.~(\ref{eq:thermal_temp})), and $T_{cat}$ (Eq.~\ref{eq:T_cat_on}) are discretized with discretization steps of $\delta t_1=1~sec$ over the short prediction horizon, and $\delta t_2=10~sec$ over the long prediction horizon.}}~In Eq. (\ref{eq:Tcat_constraint}), the minimum TWC temperature, denoted by $T_{cat}^{min}$, is defined according to the catalyst light-off temperature of $250^oC$~\cite{Amini_CCTA19,Amini_SAE}. If $T_{cat}~\ge~250^oC$, the MPC enforces this limit as a hard constraint to guarantee $T_{cat,init}$ does not drop to below the light-off temperature.~For this study, the terminal $SOC$ is allowed to vary within $\pm1\%$ of the starting SOC ($SOC_{init}$) to increase the likelihood of finding a feasible solution. To this end, the hard charge sustainability constraint in Eq. (\ref{eq:SOC_constraint}) is slightly relaxed as follows:\vspace{-4pt}
\begin{gather}\label{eq:relaxed_condition}
0.99\times SOC_{init} \le SOC_{end} \le SOC_{init}\times1.01.
\end{gather}
Once the MPC optimization problem is solved numerically, the control input at $t$ is commanded to the system, and the receding horizon is shifted by one time step ($\delta t_1$) and the long-range horizon shrinks accordingly. In this paper, we use MPCTools~\cite{risbeck2016mpctools} package for formulating the MPC optimization problem and for solving it numerically.

\vspace{-4pt}
\section{Results and Discussions}\vspace{-4pt}
Two different scenarios are considered. In the first scenario (Scenario I), eco-driving is not considered and all the vehicles are assumed to meet the human driver demand, i.e., perform normal driving. The vehicles are still connected, i.e., they broadcast their position and speed information. Furthermore, it is assumed that a high accuracy short horizon speed prediction is available to the ego-vehicle. Over the longer horizon, the data-driven algorithm shown in Figure~\ref{fig:MH_concept} is utilized. For this scenario, MPC is implemented and fuel saving benefits are evaluated. In the second scenario (Scenario II), eco-driving is considered, and the associated benefits are studied with the iPTM strategy. The eco-trajectory speed planning and long-range speed prediction are as previously discussed and shown in Figure~\ref{fig:MH_concept}. In all simulated cases, it is assumed that the engine and the TWC are warmed-up, i.e., the cold-start phase is not considered. To this end, the following initial conditions are selected: $SOC_{init}=0.6$, $T_{cl}^{min}=50^oC$, and $T_{cat}^{min}=250^oC$. \textcolor{black}{The MPC is implemented on a desktop computer with an Intel\textsuperscript{\textregistered} Core i7-8700@3.2 GHz processor. The worst case computational time in this study was $0.8~sec$ per iteration, observed at the beginning of a trip, where the overall prediction horizon is the longest.}\vspace{-4pt}

\subsection{\textbf{scenario I}: MPC-based iPTM with Normal Driving}\vspace{-4pt}
In order to evaluate the performance and demonstrate the benefits of the proposed MPC-based iPTM of connected HEVs, two ego-vehicles belonging to bin \#7 and \#8 travelling over the Plymouth Rd. corridor shown in Figure~\ref{fig:PlymouthRd_map}, are considered. {A rule-based power-split strategy with a load leveling~\cite{liu2005modeling} and additional thermal control logics is implemented as the baseline. \textcolor{black}{Particularly, the rule-based controller commands the engine to idle, if it is off, when $T_{cl}\le 50~^oC$ or $T_{cat}\le 250~^oC$. The threshold number of $T_{cl}$ is derived from the experimental data, while the thermal logic for the catalyst is adopted from literature~\cite{shahbakhti2015early}.}

\vspace{-4pt}
Three cases are considered as follows:\vspace{-8pt}
\begin{itemize}
    \item \textbf{Case A}{:~Normal driving + rule-based power-split,}\vspace{-2pt}
 \item \textbf{Case B}{:~Normal driving + MPC-based power-split with long-range accurate speed preview,}\vspace{-2pt}
  \item \textbf{Case C}{:~Normal driving + MPC-based power-split with uncertain (estimated) long-range speed preview,}
\end{itemize}
\vspace{-4pt}
In all these cases the vehicles drive normally in response to the human driver demand, i.e., their speeds are not optimized. For the MPC-based power split strategy, first, it is assumed that the entire driving cycle is known a \text{priori} (Case B). Then, this assumption is relaxed by replacing the exact speed preview with the long-term speed prediction realized through traffic data analysis (Case C). These long terms speed predictions are shown in Figures~\ref{fig:Page4_a_ND_updated}-($a$) and \ref{fig:Page5_a_ND_updated}-($a$) for ego-vehicles in bin \#7 and \#8, respectively.
%
%\vspace{-4pt}
Furthermore, in Cases B and C, $H_r=5~sec$, $\delta{t_1}=1~sec$, and $\delta{t_2}=10~sec$. Figure~\ref{fig:Results_ND_50_250_5} summarizes the fuel consumption results of these three cases for both ego-vehicles.
%\vspace{-8pt}
\begin{figure}[h!]
\begin{center}
\includegraphics[angle=0,width= \columnwidth]{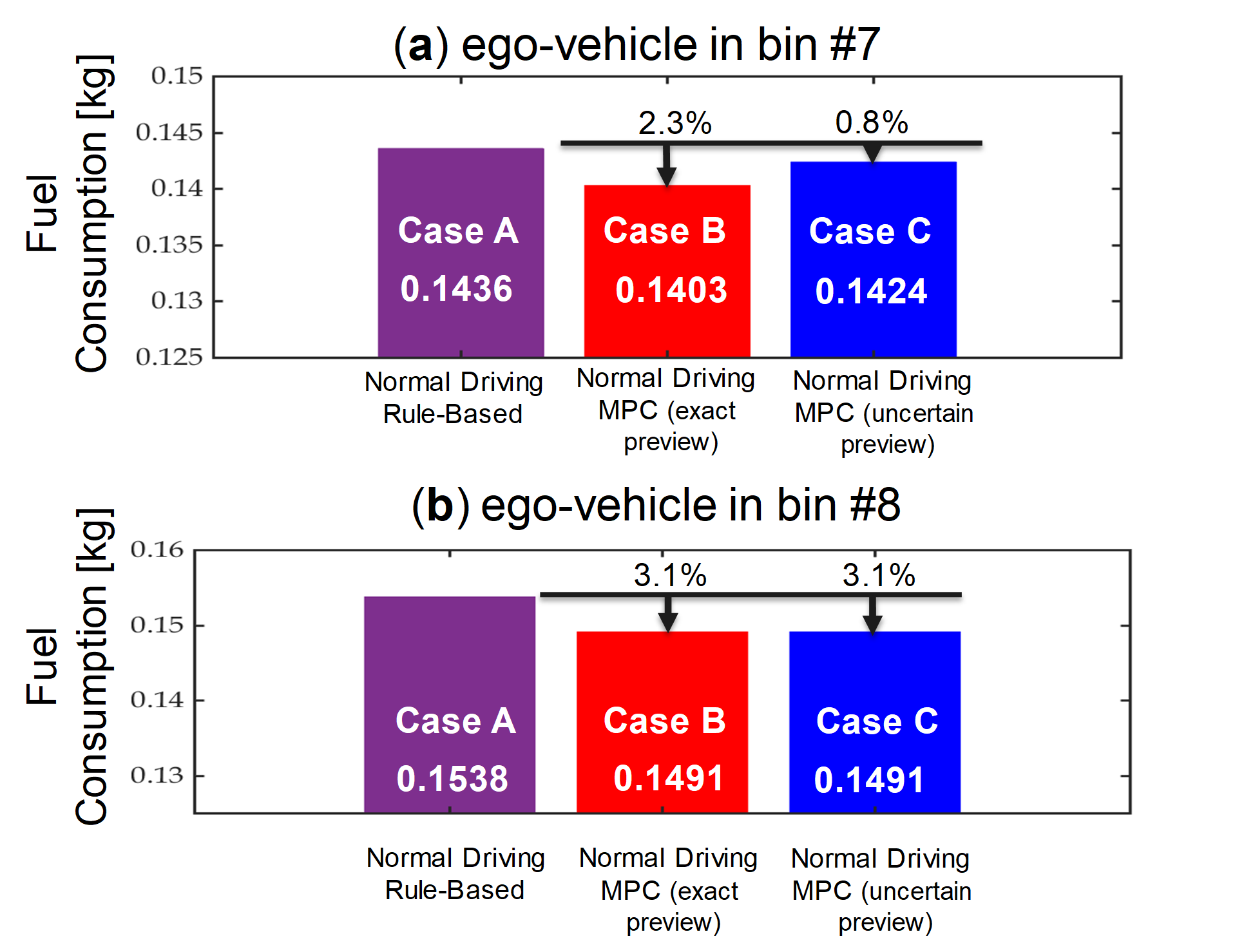}%Results_ND_50_250_5_v2_wCases 
\vspace{-0.9cm}
\textcolor{Blue}{\caption{\label{fig:Results_ND_50_250_5} The summary of fuel consumption results for (\textbf{a}) ego-vehicle in bin \#7, (\textbf{b}) ego-vehicle in bin \#8. $H_r=5~(5~sec)$, \underline{eco-driving is not considered}.}}\vspace{-2pt}
\end{center}
\end{figure}

\vspace{-4pt}
Figure~\ref{fig:Results_ND_50_250_5} shows that, compared to the rule-based controller, $2.3\%$ (ego-vehicle in bin \#7) and $3.1\%$ (ego-vehicle in bin \#8) fuel savings can be achieved using the MPC-based iPTM with exact speed preview. In the presence of speed prediction uncertainty (Case C), while the fuel saving result for the ego-vehicle in bin \#8 is not affected, it increases for the ego-vehicle in bin \#7, leading to a marginal energy saving of $0.8\%$ when comparing to Case A. This is because, as shown in Figures~\ref{fig:Page4_a_ND_updated}-($a$) and \ref{fig:Page5_a_ND_updated}-($a$), the long-term speed preview in bin \#7 is associated with larger variations, consequently affecting the MPC performance unfavorably. Note that for the ego-vehicle in bin \#7, not only the speed preview is highly uncertain, but the picked vehicle from this bin is also an outlier, see Figure~\ref{fig:Page4_a_ND_updated}-($a$).%
\vspace{-8pt}
\begin{figure}[h!]
\begin{center}
\includegraphics[angle=0,width= 0.98\columnwidth]{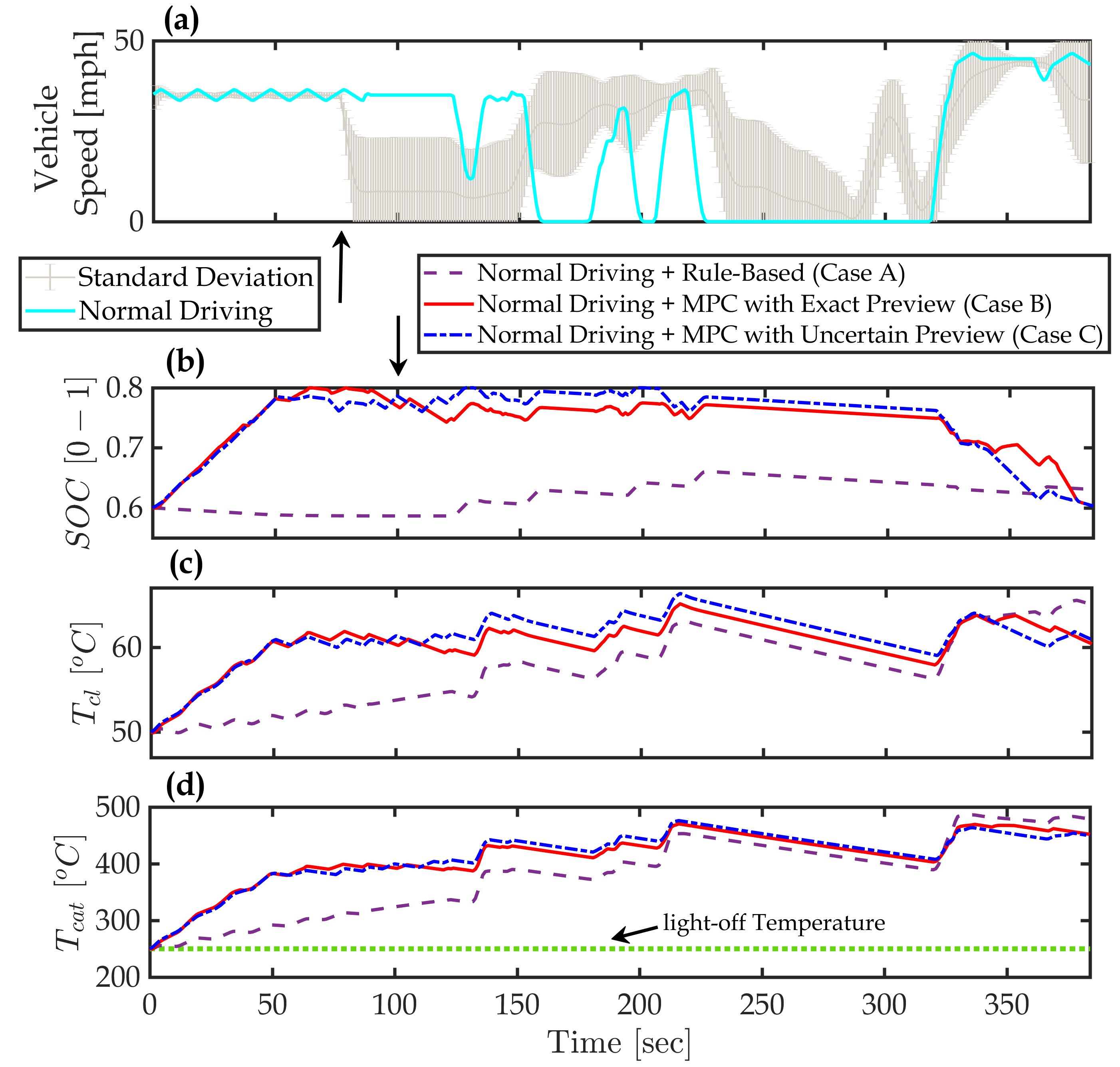} \vspace{-0.45cm}
\textcolor{Blue}{\caption{\label{fig:Page4_a_ND_updated} Power and thermal trajectories for the ego-vehicle in bin \#7: (\textbf{a}) vehicle speed, (\textbf{b}) battery $SOC$, (\textbf{c}) coolant temperature, and (\textbf{d}) catalyst temperature. $H_r=5~(5~sec)$, \underline{eco-driving is not considered}.}}\vspace{-2pt}
\end{center}
\end{figure}
\begin{figure}[h!]
\begin{center}
\includegraphics[angle=0,width= 0.98\columnwidth]{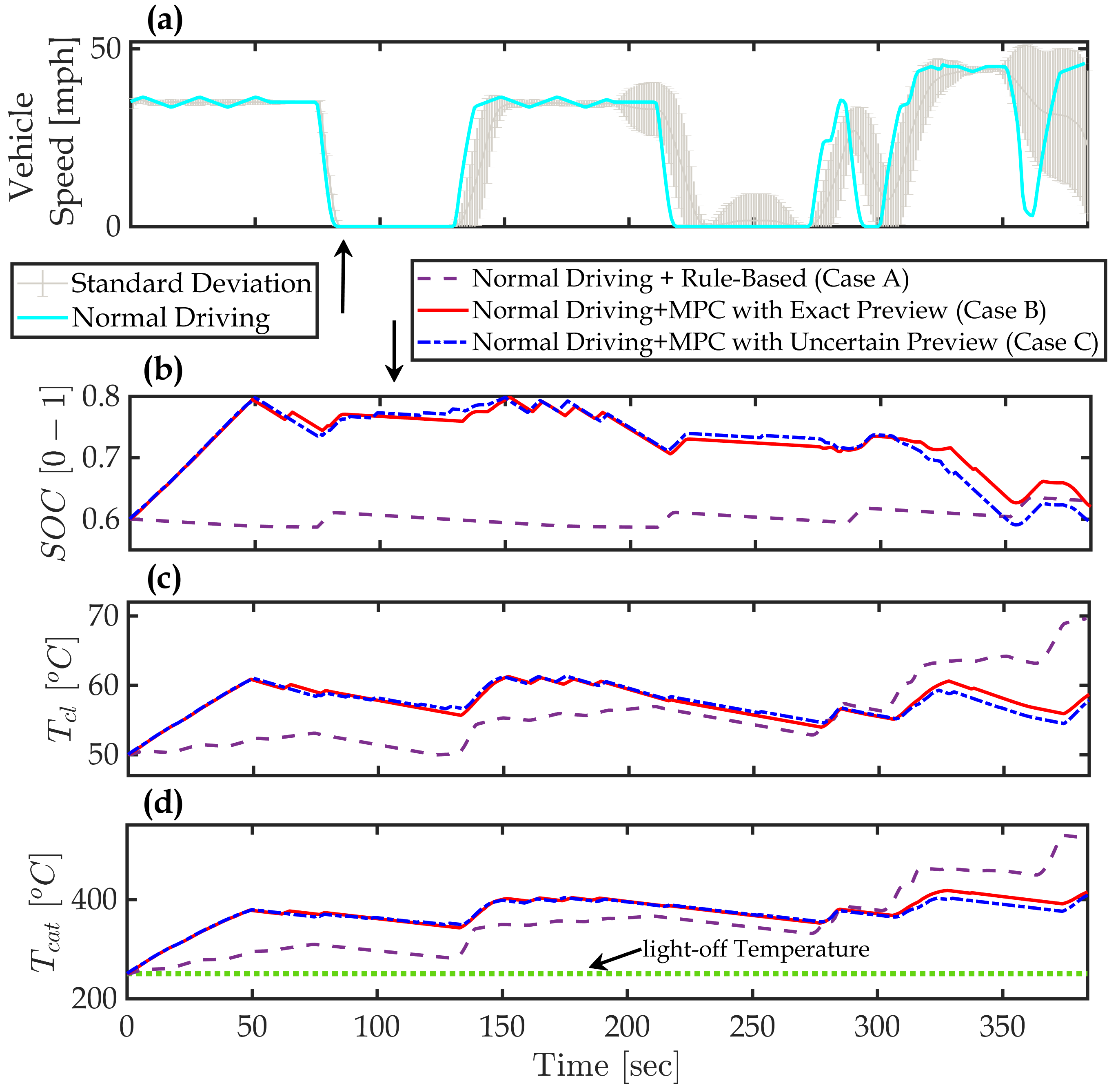} \vspace{-0.45cm}
\textcolor{Blue}{\caption{\label{fig:Page5_a_ND_updated} Power and thermal trajectories for the ego-vehicle in bin \#8: (\textbf{a}) vehicle speed, (\textbf{b}) battery $SOC$, (\textbf{c}) coolant temperature, and (\textbf{d}) catalyst temperature. $H_r=5~(5~sec)$, \underline{eco-driving is not considered}.}}\vspace{-6pt}
\end{center}
\end{figure}

\vspace{-4pt}
The power and thermal system trajectories are also shown in Figures~\ref{fig:Page4_a_ND_updated} (ego-vehicle in \#7) and \ref{fig:Page5_a_ND_updated} (ego-vehicle in \#8). Additionally, the ratio of the total engine on time to the overall trip time is summarized in Figure.~\ref{fig:Case3_EngineOnTime} for ego-vehicles in bins \#7 ($a$) and \#8 ($b$). Figures~\ref{fig:Page4_a_ND_updated}-($b$) and \ref{fig:Page5_a_ND_updated}-($b$) show that the $SOC$ trajectories with MPC for power-split is very different compared to the ones from the rule-based controller. By using MPC, during the first $50-60~sec$, the battery is charged up to its maximum limit ($0.8$). The reason for this strategy is thermal dynamics and the coupling between thermal and power systems. In particular, in order to enforce the thermal constraints and avoid engine idling during the vehicle stops, the MPC increases the coolant and catalyst temperatures at the beginning by running the engine at higher load. Since the generated engine power during this period is exceeding the traction power demand, the rest of the engine power is stored in the battery.  Note that the power trajectories from the MPC with uncertain preview are very similar to those resulted with exact speed preview. {This shows that an approximate knowledge of the long-range vehicle speed is beneficial to improve the fuel economy.}

\vspace{-4pt}
While the MPC, in all cases, has enforced the thermal constraints on $T_{cl}$ (Figures~\ref{fig:Page4_a_ND_updated}-($c$) and \ref{fig:Page5_a_ND_updated}-($c$)) and $T_{cat}$ (Figures~\ref{fig:Page4_a_ND_updated}-($d$) and \ref{fig:Page5_a_ND_updated}-($d$)), it is observed in all cases that $T_{cl}$ and $T_{cat}$ resulted by the rule-based controller are higher at the end of the driving cycle. This shows the benefit of preview information, which the MPC is able to effectively leverage by pre-heating the coolant and catalyst early in the drive.~With the energy stored in the battery, MPC allows $T_{cl}$ and $T_{cat}$ to drop within the constraints over the next segment of the trip, using less fuel to run the engine. Figure~\ref{fig:Case3_EngineOnTime} confirms this observation, where it is shown that the engine on time is reduced by up to $50\%$. Figure~\ref{fig:Case3_EngineOnTime} also shows that the engine usage for the MPC with uncertain preview (Case C) is slightly higher, as compared to MPC with exact preview (Case B). By coordinating the power and thermal systems effectively, the MPC-based iPTM approach allows the vehicle to operate in electric mode (EV) more often, specifically over the second half of the trip.\vspace{-8pt}
\begin{figure}[h!]
\begin{center}
\includegraphics[angle=0,width= 0.92\columnwidth]{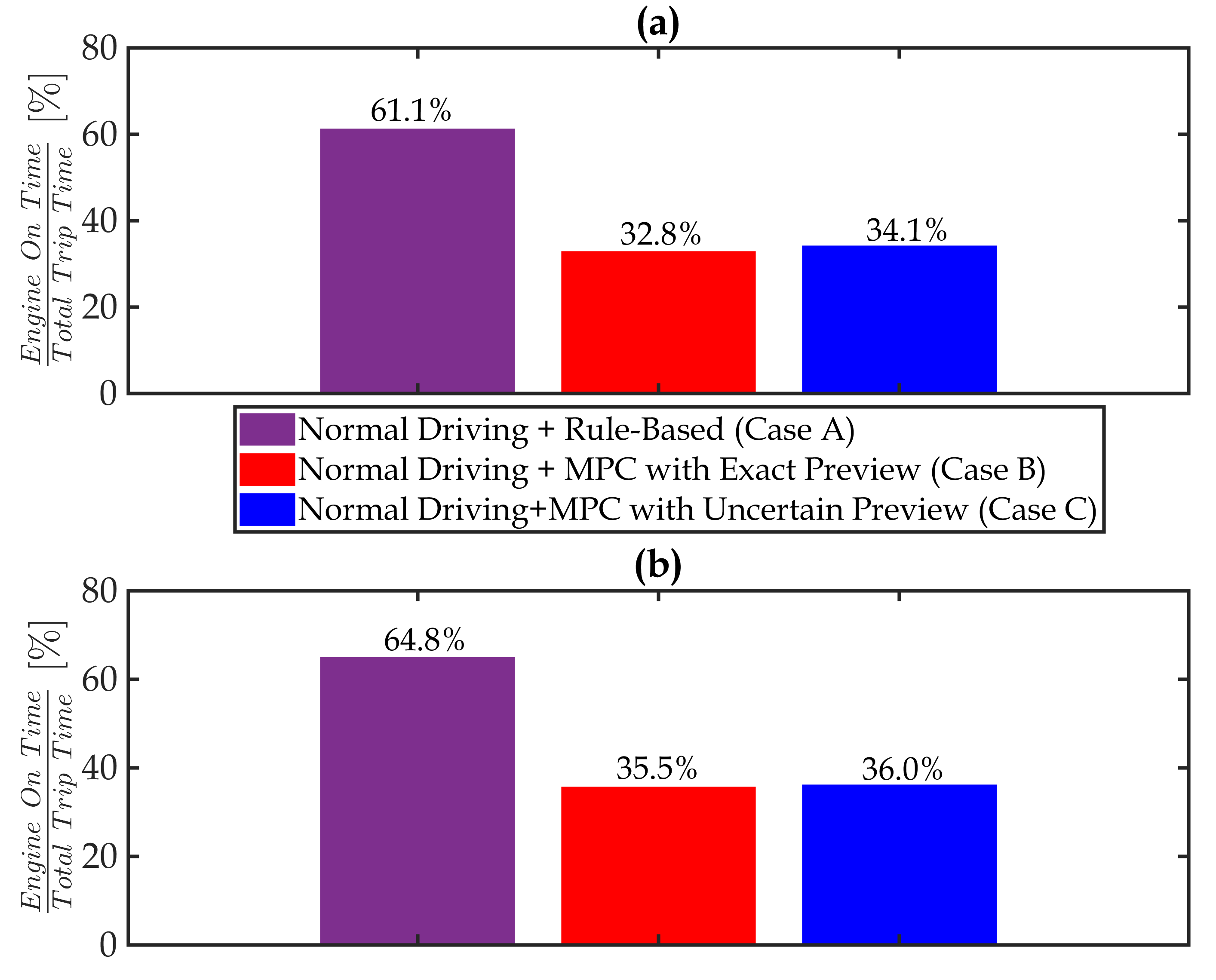} \vspace{-0.55cm}
\textcolor{Blue}{\caption{\label{fig:Case3_EngineOnTime} The ratio of the total engine on time to the overall trip time for (\textbf{a}) ego-vehicle in bin \#7, (\textbf{b}) ego-vehicle in bin \#8. $H_r=5~(5~sec)$, \underline{eco-driving is not considered}.}}\vspace{-6pt}
\end{center}
\end{figure}

\vspace{-2pt}
\subsection{\textbf{scenario II}: MPC-based iPTM with Eco-Driving}\vspace{-8pt}
The benefits of applying the proposed iPTM strategy was studied in the previous section. In this section, eco-driving is also considered, and the associated benefits will be evaluated sequentially along with the MPC-based power-split approach. To this end, four cases are considered as follows:\vspace{-12pt}
\begin{itemize}
    \item \textbf{Case I}{:~Normal driving + rule-based power-split,}\vspace{-2pt}
    \item \textbf{Case II}{:~Eco-driving + rule-based power-split,}\vspace{-2pt}
 \item \textbf{Case III}{:~Eco-driving + MPC-based power-split with long-range accurate speed preview,}\vspace{-2pt}
  \item \textbf{Case IV}{:~Eco-driving + MPC-based power-split with uncertain (estimated) long-range speed preview,}
\end{itemize}

\vspace{-6pt}
In Case II, while the power-split logic is the same as the rule-based one used in Case I, both ego-vehicles follow the planned eco-trajectories (i.e., they perform eco-driving) based on the strategy described earlier in the speed planning and prediction section. In Case III and IV, similar to Case II, both ego-vehicles implement eco-driving and the power-split logic is also based on the MPC presented in Eq.~(\ref{eq:variable_timescale_MPC_formulation}). The receding horizon $H_r$ and update rate $\delta{t_1}$ and $\delta{t_2}$ are the same as Scenario I.

\vspace{-4pt}
The fuel consumption results of these four cases for the two considered ego-vehicles are summarized in Figure~\ref{fig:Results_50_250_5}. By comparing Case I to II, it can be seen that a significant ($9.5\%$ for the vehicle in bin \#7 and $10.3\%$ for the one in bin \#8) reduction on the fuel consumption can be achieved through eco-driving. Once the MPC-based iPTM strategy is implemented (Case III), Figure~\ref{fig:Results_50_250_5} shows that the fuel consumption is further reduced by $2.4\%$ on average, as compared to Case II. These fuel saving results, however, are not realistic as Case III assumes the long-range speed previews are accurate. Once the ideal speed previews are replaced with those estimated ones (Case IV), an increase in the fuel consumption is observed for both ego-vehicles compared to Case III. For the ego-vehicle in bin \#7, the fuel consumption of Case IV increases by $1.5\%$ comparing to Case III. On the other had, this increase in the fuel consumption for the second ego-vehicle in bin \#8 in only $0.3\%$. The different impact of the speed uncertainty on the fuel economy can be explained according to the speed variations in bins \#7 and \#8 as shown in Figure~\ref{fig:FinalFig_1to10_withAggregated}. The speed trajectories of these ego-vehicles are also shown in Figures.~\ref{fig:Page4_a_updated}-($a$) and \ref{fig:Page5_a_updated}-($a$). As %\{can be seen from these figures, since }
the uncertainty in the long-range vehicle speed prediction is higher for the ego-vehicle in bin \#7, as compared to the ego-vehicle in bin \#8, the uncertainty impact on the fuel economy is more pronounced in bin \#7~%{for the first ego-vehicle }
when comparing Cases III and IV.
\vspace{-10pt}
\begin{figure}[h!]
\begin{center}
\includegraphics[angle=0,width= \columnwidth]{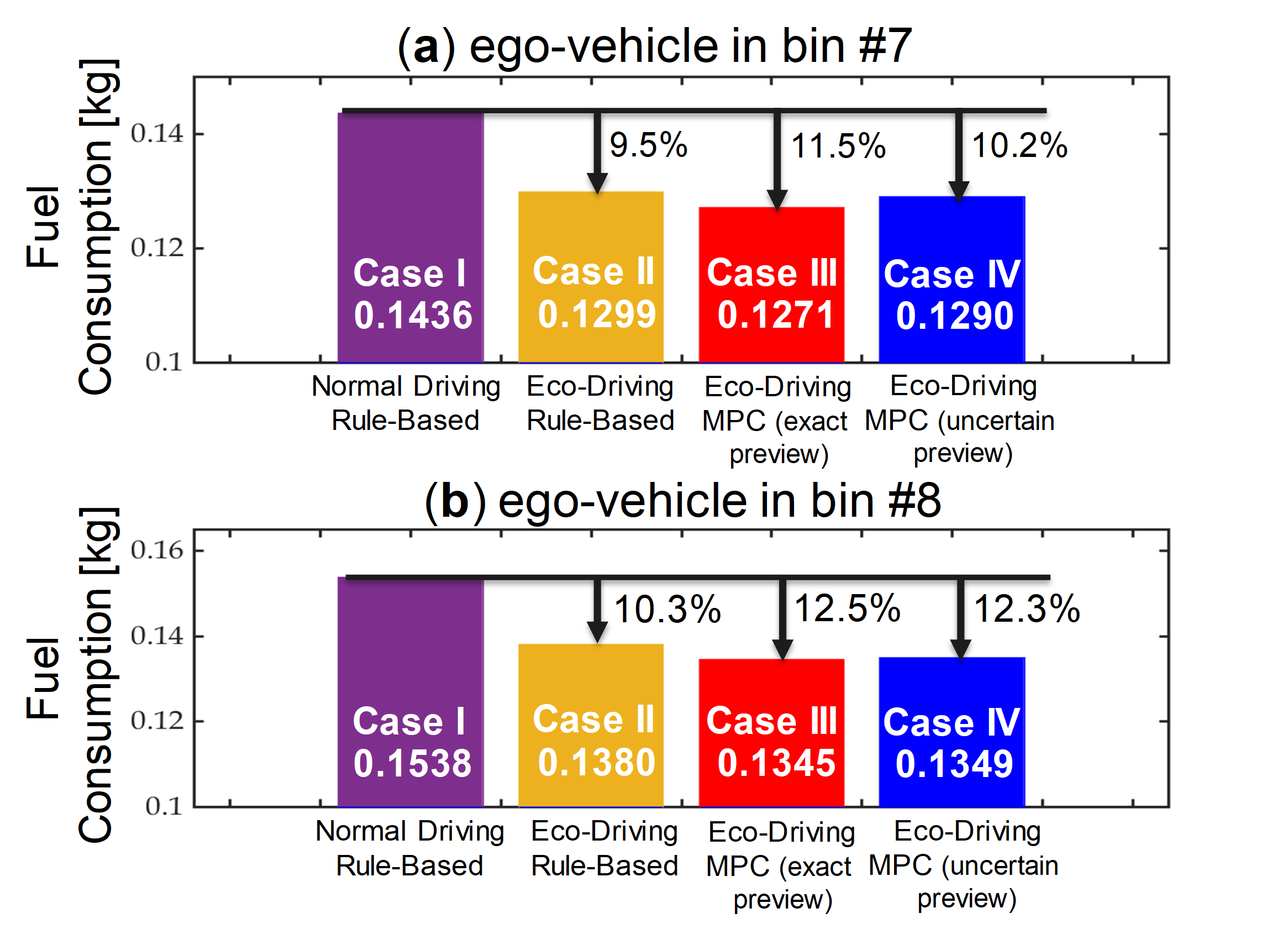} \vspace{-0.75cm}
\textcolor{Blue}{\caption{\label{fig:Results_50_250_5} The summary of fuel consumption results from Cases I to IV for (\textbf{a}) ego-vehicle in bin \#7, (\textbf{b}) ego-vehicle in bin \#8. $H_r=5~(5~sec)$.}}\vspace{-4pt}
\end{center}
\end{figure}
\begin{figure}[h!]
\begin{center}
\includegraphics[angle=0,width= 0.98\columnwidth]{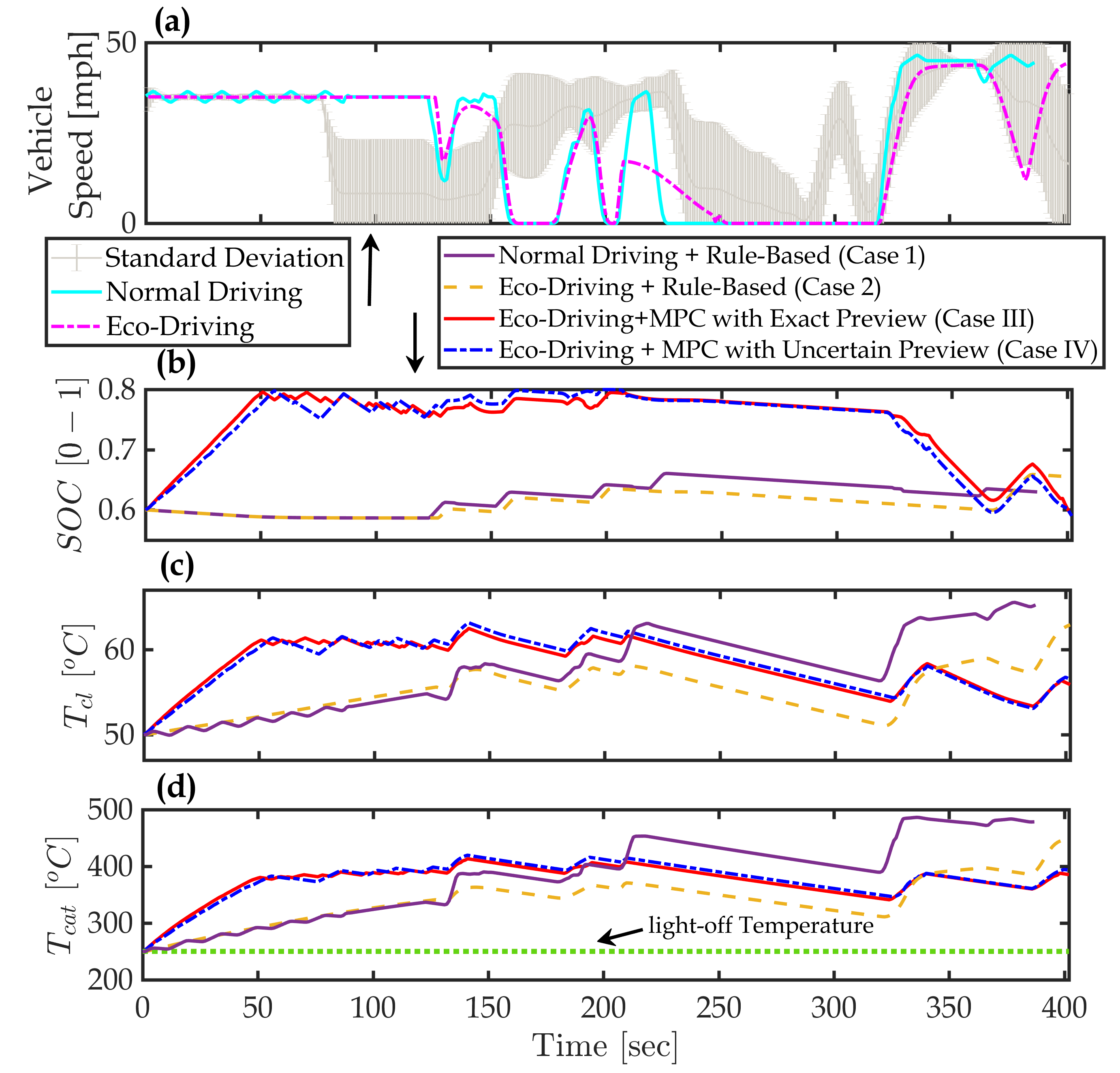} \vspace{-0.45cm}
\textcolor{Blue}{\caption{\label{fig:Page4_a_updated} Power and thermal trajectories for the ego-vehicle in bin \#7: (\textbf{a}) vehicle speed, (\textbf{b}) battery $SOC$, (\textbf{c}) coolant temperature, and (\textbf{d}) catalyst temperature. $H_r=5~(5~sec)$.}}\vspace{-6pt}
\end{center}
\end{figure}
\begin{figure}[h!]
\begin{center}
\includegraphics[angle=0,width= 0.98\columnwidth]{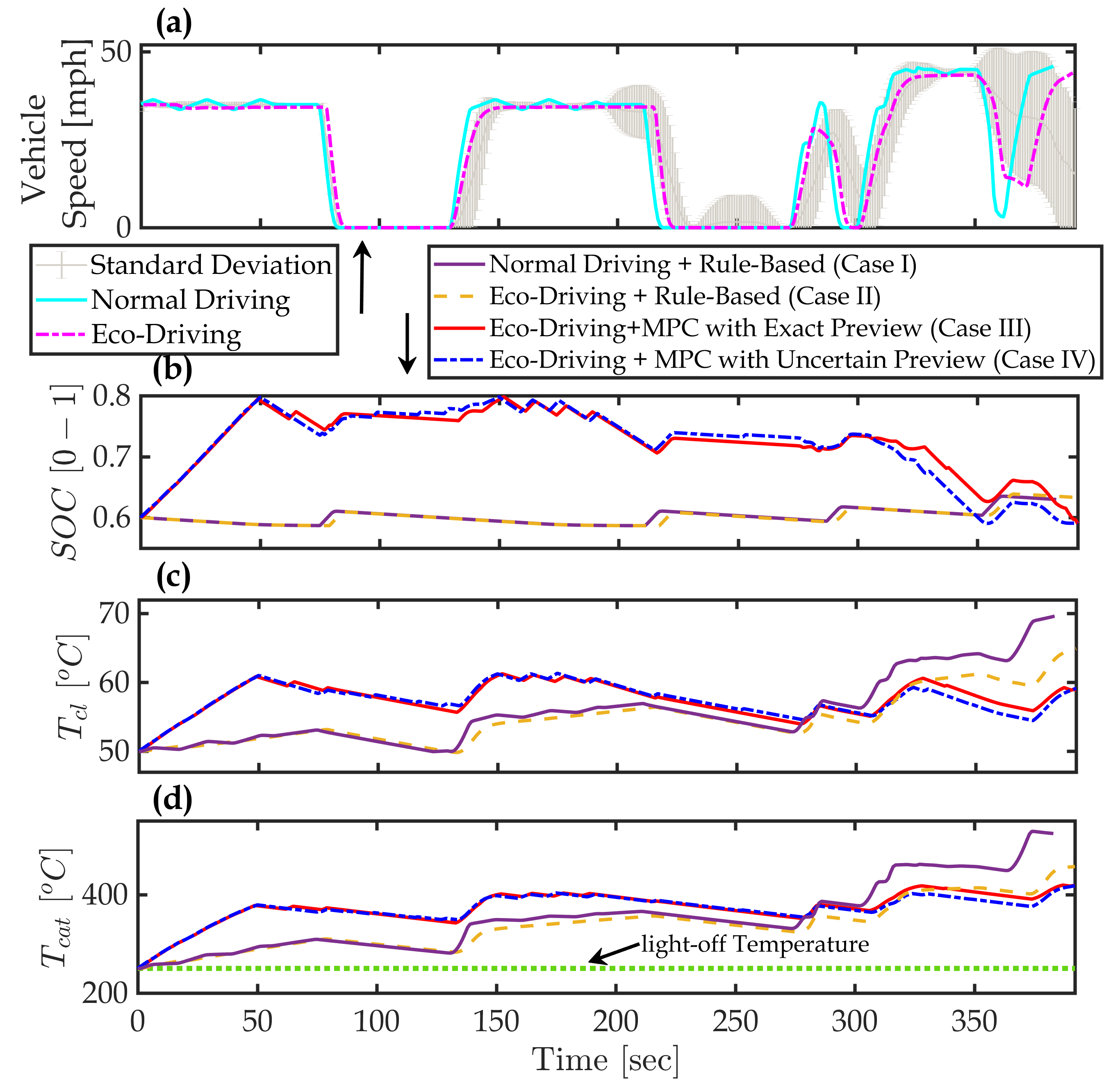} \vspace{-0.45cm}
\textcolor{Blue}{\caption{\label{fig:Page5_a_updated} Power and thermal trajectories for the ego-vehicle in bin \#8: (\textbf{a}) vehicle speed, (\textbf{b}) battery $SOC$, (\textbf{c}) coolant temperature, and (\textbf{d}) catalyst temperature. $H_r=5~(5~sec)$.}}\vspace{-10pt}
\end{center}
\end{figure}

\vspace{-4pt}
Figures~\ref{fig:Page4_a_updated} and ~\ref{fig:Page5_a_updated} present the power and thermal trajectories of Cases I to IV for ego-vehicle in bins \#7 and \#8, respectively. By comparing the normal driving and eco-driving speed trajectories in Figures~\ref{fig:Page4_a_updated}-($a$) and ~\ref{fig:Page5_a_updated}-($a$), one can see how the eco-trajectory speed planning strategy modifies the ego-vehicles' speed profiles. Compared to normal driving, the eco-vehicles have a smoother speed profiles with less aggressive acceleration events, see~\cite{AminiACC19,yang2019eco} for further details on the impacts of the incorporated eco-driving strategy. With $T_{cl,init}=50^oC$ and $T_{cat,init}=250^oC$ (above light-off), it can be observed that for all cases $T_{cat}$ is maintained above the light-off temperature (Figures~\ref{fig:Page4_a_updated}-($d$) and ~\ref{fig:Page5_a_updated}-($d$)). A similar observation can be made for $T_{cl}$, see Figures~\ref{fig:Page4_a_updated}-($c$) and ~\ref{fig:Page5_a_updated}-($c$). %\vspace{-4pt}

\vspace{-4pt}
The most notable difference in the ego-vehicles trajectories is the $SOC$ (Figures~\ref{fig:Page4_a_updated}-($b$) and ~\ref{fig:Page5_a_updated}-($b$)) from Cases III and IV, as compared to the ones from Cases I and II. Due to the different battery charge/discharge strategy from the MPC-based approach, which means different operating conditions for the engine, the thermal responses ($T_{cat},T_{cl}$) are also different. Compared to the rule-based power-split logic, $SOC$ of MPC-based logic varies in a much broader range, meaning the battery is used in a more aggressive way. For both ego-vehicles, $SOC$ rises to its upper constraint ($0.8$) in the first $50~sec$ of the trip, while $T_{cl}$ rises to about 60$~^oC$.The low temperature degradation of the engine performance disappears when the coolant temperature reaches to around 60$~^oC$~\cite{kim2016thermal,gong2019integrated}. The MPC's knowledge of trip length and projected conditions allow it to drive $T_{cl}$ to the higher efficiency region faster than the rule-based controller that is designed to operate without preview information.%To this end, the MPC puts efforts to rise $T_{cl}$ to the efficient region fast. The rule-based controller, on the other hand, does not leverage the sensitivity of the fuel consumption to the coolant temperature and only tried to maintain $T_{cl}$ above $50^oC$. 
~Moreover, with battery charged sufficiently and coolant at high temperature, enough electrical and thermal energies are stored for traction and heating demands. Thereby, the MPC-based iPTM strategy lets the vehicle operate in the EV mode over the second half of the trip. Since, the speed preview is not leveraged for Cases I and II with the rule-based power-split controller, the engine and the battery are used less efficiently, leading to higher coolant and TWC temperatures at the end of the trip. This means that extra thermal energy is generated by the engine in Cases I and II, which are wasted eventually.%Both $SOC$ and $T_{cl}$ decreases in this period

\vspace{-4pt}
The ratio of the total engine on time to the overall trip time is calculated from Cases I to IV, and the results are shown in Figure.~\ref{fig:Case1_EngineOnTime}. It can be seen that compared with Cases I and II, the proposed iPTM strategy (Cases III and IV) cuts the engine usage by half, approximately, even with an estimated long-range speed preview (Case IV).% uses engine less often, and fully utmilizes the high efficiency of the electrical mode.
\vspace{-10pt}
\begin{figure}[h!]
\begin{center}
\includegraphics[angle=0,width= 0.92\columnwidth]{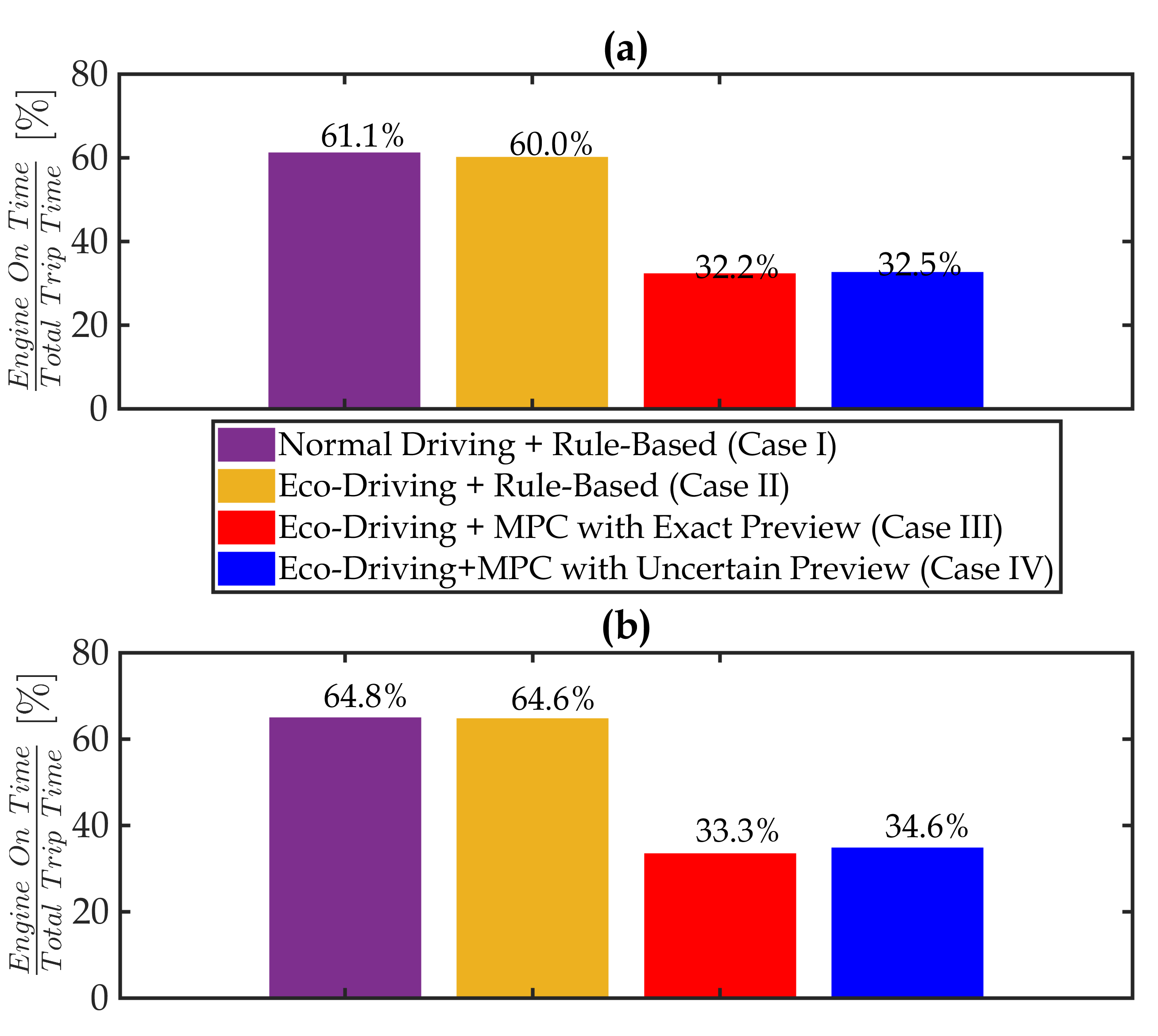} \vspace{-0.55cm}
\textcolor{Blue}{\caption{\label{fig:Case1_EngineOnTime} The ratio of the total engine on time to the overall trip time for (\textbf{a}) ego-vehicle in bin \#7, (\textbf{b}) ego-vehicle in bin \#8. $H_r=5~(5~sec)$.}}\vspace{-4pt}
\end{center}
\end{figure}

\vspace{-4pt}
\subsection{Impact of Short-Range Prediction Horizon on the Fuel Economy}\vspace{-4pt}
As discussed in the previous sections, the long-range speed prediction uncertainty has a direct impact on the MPC-based iPTM results according to Figures~\ref{fig:Results_ND_50_250_5} and~\ref{fig:Results_50_250_5}. This impact is larger especially for the ego-vehicle in bin \#7 with higher speed variations (Figure~\ref{fig:FinalFig_1to10_withAggregated}). As can be seen from Figure~\ref{fig:Results_50_250_5}-($a$), Case IV leads to a fuel consumption that is higher than Case III by $1.5\%$, and only slightly lower than Case II by $0.7\%$, meaning that the uncertainty in the long-range speed significantly degrades the MPC performance. In this section, we focus on the short-range prediction horizon and investigate the impact of this moving horizon length ($H_r$) on the fuel economy results of the MPC strategy. 

\vspace{-4pt}
Figure.~\ref{fig:Results_H_r_5_20} shows the results of increasing $H_r$ from $5~(5~sec)$ to $20~(20~sec)$ for Cases III and IV while all other parameters remain the same, i.e., $T_{cl,init}=50^oC$, $T_{cat,init}=250^oC$. Note that over the short-range moving horizon $H_r$, it is still assumed that the vehicle speed can be predicted accurately. As a reminder, in Case III, the entire driving cycle is assumed to be known a \textit{priori}. As shown in Figure.~\ref{fig:Results_H_r_5_20}, extending $H_r$ further improves the fuel economy of Case IV, and delivers results close to Case III (i.e., the benchmark case). This observation applied even to the ego-vehicle in bin \#7, which is imposed to large long-range speed prediction uncertainty.
\vspace{-6pt}
\begin{figure}[h!]
\begin{center}
\includegraphics[angle=0,width= \columnwidth]{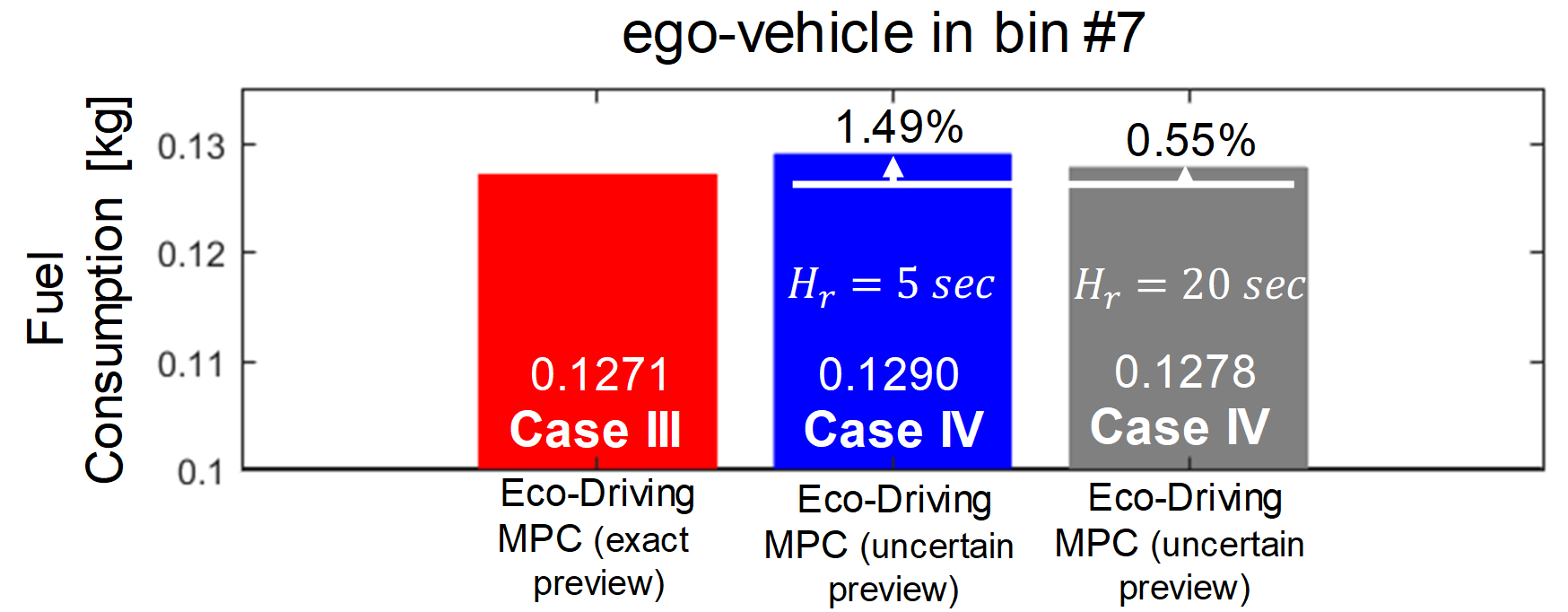} \vspace{-0.75cm}
\textcolor{Blue}{\caption{\label{fig:Results_H_r_5_20} The impact of extending $H_r$ on the MPC-based iPTM results with $T_{cl,init}=50^oC$, $T_{cat,init}=250^oC$.}}\vspace{-4pt}
\end{center}
\end{figure}

\vspace{-4pt}
The MPC-based iPTM approach leverages the speed preview over both short and long prediction horizons to fully utilize the battery and optimize the thermal responses of the engine and the catalytic converter. While a relatively long prediction horizon is required for planning the thermal and energy states ($SOC,~T_{cl},~T_{cat}$), our analysis shows that an approximate estimation of the long-range vehicle speed can serve the purpose. At the same time, Figure~\ref{fig:Results_H_r_5_20} shows that extending the high-resolution short prediction horizon ($H_r$) can further enhance the fuel economy of the connected HEVs. This observation can be explained according the the specific information available over $H_r$ at faster update rate of $\delta t_1=1~sec$. The MPC-based power-split logic drives the battery $SOC$ to its upper constraint. This $SOC$ upper constraint, in this paper, is enforced as a hard constraint, meaning its violation is not tolerated. While $SOC$ is evolving close to its limit with a short $H_r$, if any braking event occurs, the battery is not able to recapture this free energy through regenerative braking. This is because $SOC$ is already at its limit and any battery charging exceeding the battery power demand leads to $SOC$ upper constraint violations. With a longer $H_r$, however, the battery has enough lead time to discharge early on and create enough charging capacity prior to the breaking event. 

\vspace{-4pt}
Overall, for the Plymouth Rd. corridor shown in Figure~\ref{fig:PlymouthRd_map}, the MPC-based iPTM divides the trip into three stages. In the first stage, the $SOC$ reaches its upper constraint, and $T_{cl}$ and $T_{cat}$ increase to improve the engine efficiency.~In the second stage, $SOC$ evolves close to its upper constraint. Finally, in the third stage and according to the acquired knowledge from the historic data about the rest of the trip, vehicle drives primarily in EV mode as $SOC$, $T_{cl}$ and $T_{cat}$ decrease within the constraints. Such an iPTM strategy, combined with eco-driving which mainly reduces the traction power losses, results in $11-12\%$ fuel saving, on average, compared with the baseline case with normal driving and rule-based power-split controller.

\section{Summary and Conclusions}\label{Stage_II_Conclusion}\vspace{-4pt}
We presented an optimization framework for integrated power and thermal management (iPTM) of connected HEVs based on model predictive control (MPC). The proposed iPTM strategy is based on a multi-range vehicle speed prediction and planning scheme, which includes short- and long-range speed previews. The short-range speed prediction is realized through V2X-based eco-trajectory speed planning for connected vehicles (CVs) for eco-driving. The long-range speed preview, on the other had, is estimated by analyzing the traffic data collected from the CVs travelling over the same corridor as the target ego-vehicle. This multi-range speed prediction strategy was applied to an arterial corridor in Ann Arbor, MI, modelled in VISSIM. In order to evaluate the performance of the MPC-based iPTM strategy, a power-split HEV model, with experimentally validated battery $SOC$, engine coolant temperature, and three-way catalytic converted temperature models, was used.

\vspace{-4pt}
The results of implementing the iPTM framework combined with eco-driving, in comparison with a baseline scenario with a rule-based power-split logic, showed that:\vspace{-12pt}
\begin{itemize}
    \item assuming the entire driving cycle is known a \textit{priori}, the optimization approach for power-split control provides an energy saving of up to $2.7\%$ (without eco-driving) and $2.4\%$ (with eco-driving). %Notably, the benefits of the proposed iPTM strategy are more pronounced for the cold-start case with more aggressive enforcement of therm system constraints.
    \vspace{-4pt}
    \item eco-driving reduces fuel consumption by $9.5-10.3\%$ through vehicle speed optimization and reducing traction power losses.\vspace{-4pt}
    \item the uncertainty in long-term speed prediction can significantly impact the performance of the iPTM. The data mining approached helped to generate long-range speed preview with relatively low variations, resulting in a robust MPC performance. For those cases with uncertain preview, the MPC showed only marginal improvements.\vspace{-4pt}
    \item the iPTM strategy coordinated the power and thermal loops efficiently, leading to a major decrease in the engine use time by up to $50\%$. %(warmed-up scenario) and $60\%$ (cold-start scenario).
    \vspace{-4pt}
\end{itemize}
Future works will focus on improving the robustness of the iPTM strategy, and applying advanced data classification algorithms to further enhance the long-range vehicle speed prediction accuracy. \textcolor{black}{This process will involve studying iPTM performance over drive cycles with longer distances. Moreover, the cold-start operation of the engine, which can significantly alter the powertrain behaviours, was not considered in this study. In our future investigations, we will consider the engine cold start phase with an enhanced model which is able to accommodate cold-start characteristic.}

%%%%%%%%%%%%%%%%%%%%%%%%%%%%%%%%%%%%%%%%%%%%%%%%%%%%%%%
%%%%%%%%%%%%%%%
\section{Contact Information}\vspace{-4pt}
{Qiuhao Hu}, \underline{qhhu@umich.edu} 

\vspace{-4pt}
{Mohammad Reza Amini}, \underline{mamini@umich.edu} 

\vspace{-4pt}
{Jing Sun}, \underline{jingsun@umich.edu} 

%\vspace{-4pt}
%{Julia Buckland Seeds}, \underline{jingsun@umich.edu}

%%%%%%%%%%%%%%%%%%%%%%%%%%%%%%%%%%%%%%%%%%%%%%%%%%%%%%%%%%%%%%%%%%%%%%
%
%%%%%%%%%%%%%%%%%%%%%%%%%%%%%%%%%%%%%%%%%%%%%%%%%%%%%%%%%%%%%%%%%%%%%%
\section*{Acknowledgments}\vspace{-4pt}
The University of Michigan portion of this work is funded in part by the United States Department of Energy (DOE), ARPA-E NEXTCAR program under award No. DE-AR0000797.
%%%%%%%%%%%%%%%%%%%%%%%%%%%%%%%%%%%%%%%%%%%%%%%%%%%%%%%%%%%%%%%%%%%%%%
%\newpage
\bibliographystyle{unsrt}
\bibliography{SAE2020_bib}

\onecolumn

\section*{Appendix}
\vspace{-8pt}
\emph{\textbf{Nomenclature}}:\\
\vspace{-10pt}
%\section*{Nomenclature}
%\vspace{-0.7cm}
\begin{table}[h]
\addvspace{0.2in}
\renewcommand\arraystretch{1.2}
\begin{tabular}{l l}
%\vspace{-0.16cm}
$H_r$ &Short receding horizon, [$step$]\\ 
$H_s$ &Long shrinking horizon, [$step$]\\ 
$\dot{m}_{fuel}$ & fuel flow rate, [$kg/sec$]\\ 
$P_{bat}$ &Battery power, [$W$]\\
$P_{eng}$ &Engine power, [$W$]\\
%\end{tabular}
%
%\section*{}
%\vspace{-0.35cm}
%\begin{tabular}{l l}
%
$P_{HVAC}$ &HVAC system power, [$W$]\\ 
$P_{pump}$ &Coolant pump power, [$W$]\\
$P_{d}$ &Traction power demand, [$W$]\\ 
$\dot{Q}_{air}$ &rate of the heat rejected by air convection, [$W$]\\ 
$\dot{Q}_{exh}$ &rate of heat rejected in the exhaust, [$W$]\\ 
$\dot{Q}_{fuel}$ &heat release rate in the combustion process, [$W$]\\ 
$\dot{Q}_{heat}$ &rate of heat exchanged for cabin heating, [$W$]\\ 
$R_{int}$ &Battery resistance, [$\Omega$]\\ 
$SOC$ &Battery state-of-charge, [$0-1$]\\ 
$T_{amb}$ &Ambient temperature, [$^oC$]\\
$T_{cat}$ &Catalyst temperature, [$^oC$]\\
$T_{cl}$ &Engine coolant temperature, [$^oC$]\\
$U_{oc}$ &Battery open circuit voltage, [$V$]\\ 
$V_{veh}$ &Vehicle speed, [$m/sec$]\\ 
$\delta t_1,~\delta t_2$ &sampling times, [$sec$]\\
%$\delta t_2$ &Slow sampling time, [$sec$]\\
$\omega_e$ &Engine speed, [$rad/sec$]\\
$\tau_e$ &Engine torque, [$N.m$]\\
\vspace{-0.16cm}
\end{tabular}
\end{table}

\vspace{-15pt}
\emph{\textbf{Acronyms}}:\\
\vspace{-10pt}
%\section*{Nomenclature}
%\vspace{-0.7cm}
\begin{table}[h]
\addvspace{0.2in}
\renewcommand\arraystretch{1.2}
\begin{tabular}{l l}
%\vspace{-0.16cm}
\textit{CAV} & Connected and automated vehicle \\
\textit{CV} & Connected vehicle \\
%\textit{DP} & Dynamic programming \\
\textit{EV} & Electric vehicle\\
\textit{iPTM} &Integrated power and thermal management\\ 
\textit{HEV} & Hybrid electric vehicle\\
\textit{HVAC} & Heating, ventilation and air conditioning\\
\textit{MPC} & Model predictive control \\
%\textit{PMP} & Pontryagin’s  Maximum  Principle \\
\textit{TWC} & Three-way catalyst \\
\textit{V2I} & Vehicle to infrastructure \\
\textit{V2V} & Vehicle to Vehicle \\
\vspace{-0.16cm}
\end{tabular}
\end{table}

\vspace{-10pt}
\emph{\textbf{Parameters of $T_{cl}$ model (Eq.~(\ref{eq:T_cat_on}))}}:%\\
\vspace{-8pt}

$\alpha_1=-1.6065e-2$, $\alpha_2=-1.8535e-06$, $\alpha_3=9.8852e-3$, $\alpha_4=-8.2564e-05$, $\alpha_5=5.1029e-3$, $\alpha_6=-1.6444e-4$,
$\alpha_7=1.5473e-6$, $\alpha_8=6.8078$

$\beta_1=-1.00e-3$, $\beta_2=-0.200$

The unit of the variables in (Eq.~(\ref{eq:T_cat_on})) are:\\
$V_{veh}~[m/s]$, $\omega_{e}~[rad/s]$, $\tau_{e}~[Nm]$, $T_{cat}~[^oC]$, $T_{amb}~[^oC]$

\end{document}